\documentclass[fleqn,usenatbib]{mnras}

\usepackage{newtxtext,newtxmath}

\usepackage[T1]{fontenc}
\usepackage{orcidlink}

\DeclareRobustCommand{\VAN}[3]{#2}
\let\VANthebibliography\thebibliography
\def\thebibliography{\DeclareRobustCommand{\VAN}[3]{##3}\VANthebibliography}

\newcommand{\logRHK}{log\,$R'_\mathrm{HK}$}

\newcommand{\feat}{activity-sensitive features}
\newcommand{\nstar}{14}

\usepackage{booktabs}

\usepackage{graphicx}	
\usepackage{amsmath}	

\usepackage{nccmath}

\title[SRA for astrophysical noise mitigation]{Spectral Ratio Analysis: probing of a new suite of stellar activity indicators as a tool for astrophysical noise mitigation}

\author[J. C. Costes et al.]{
\parbox{\textwidth}{
Jean C. Costes$^{1}$\thanks{E-mail: jcostes01@qub.ac.uk},
Christopher A. Watson$^{1}$\orcidlink{0000-0002-9718-3266},
Ernst de Mooij$^{1}$,
Katlyn L. Hobbs$^{1}$,
Dana Clarice S. Yaptangco$^{2}$,
Yvonne C. Unruh$^{2}$,
Megan Bedell$^{3}$,
Nadège Meunier$^{4}$,
Thomas D. Mitchell$^{1,5}$
}
\vspace{0.3cm}
\\
$^{1}$Astrophysics Research Centre, School of Mathematics and Physics, Queen’s University Belfast, Belfast BT7 1NN, UK\\
$^{2}$Department of Physics, Imperial College London, London SW7 2AZ, UK\\
$^{3}$Center for Computational Astrophysics, Flatiron Institute, 162 5th Avenue, New York, NY 10010, USA\\
$^{4}$Univ. Grenoble Alpes, CNRS, IPAG, 38000 Grenoble, France\\
$^{5}$ Armagh Observatory \& Planetarium, College Hill, Armagh, BT61 9DG, UK\\
}

\date{Accepted XXX. Received YYY; in original form ZZZ}

\pubyear{\the\year{}}

\begin{document}
\label{firstpage}
\pagerange{\pageref{firstpage}--\pageref{lastpage}}
\maketitle

\begin{abstract}

Stellar activity is the main barrier to detecting and/or confirming low-mass/long-period (and Earth-analogue) planets using radial-velocity (RV) measurements. Searching for reliable indicators that better trace magnetic activity may be key for both distinguishing more clearly between stellar and planetary signals, and for probing the underlying physics occurring on the stellar surface. In this work we have compared observations taken for magnetically active and inactive stellar phases over multiple time scales to study the spectral imprint due to varying stellar activity. This serves as a proof-of-concept demonstration of a technique (named Spectral Ratio Analysis, SRA) that can be used to isolate activity-driven changes directly in the stellar photospheric absorption lines where RVs are measured. Using \nstar~relatively quiet and well sampled G- and K-type stars that show stellar activity cycles, we identified hundreds of activity-sensitive spectral features. Reducing this variability information into two global metrics -- amplitude and velocity shift -- uncovers potential evidence of a decoupling of the photospheric and chromospheric responses to stellar activity in earlier-type stars. Additionally, potential signatures of the variations in the magnitude of the suppression of the convective blueshift throughout the activity cycle were observed via SRA. Finally, we show that these SRA indicators better capture RV variability than classical activity proxies, such as the chromospheric \logRHK\ index and other cross-correlation function-based parameters such as BIS and FWHM, by up to a factor of two. The direct link between photospheric line behaviour and stellar-induced RV variability offers a promising path for improving astrophysical noise mitigation.

\end{abstract}


\section{Introduction}
While thousands of exoplanets have been discovered in the last 30 years, the discovery of a planet truly resembling the Earth has remained elusive. Upcoming flagship missions aim to change this. For example, the European Space Agency mission PLATO (PLAnetary Transits and Oscillations), scheduled for launch in early 2027, is specifically tasked with detecting the transits of Earth-analogue planets. Uniquely, for a space-based transit mission, PLATO includes (as a deliverable - see \citealt{Rauer2025}) the measurements of the masses of terrestrial planets located in the habitable zones of F5V to K7V stars. Currently, extreme precision radial-velocity (RV) ground-based surveys are required in order to provide the masses of these potential “Earth-twin” planets. However, as RV measurements become more precise thanks to the latest highly wavelength-stabilised spectrographs (e.g. ESPRESSO, see \citealt{Pepe2021}) they also become progressively more sensitive to RV variability driven by stellar noise. It is these signals that remain a significant barrier towards the ultimate goal of discovering Earth-like planets.

Main sequence stars, such as the Sun, generate magnetic activity that causes variations in the shape of the stellar spectral lines, which in turn induces apparent RV signals. In particular, the presence of faculae (or plage), which suppresses the underlying convection in the stellar photosphere, dominates the stellar induced RVs in low-activity stars like the Sun (e.g. \citealt{Haywood2016, Milbourne2019, Meunier2024}) and is thought to be one of the largest impediments to detecting Earth-like planets. However, it is difficult to track faculae in observations due to their relatively high filling factors, their more homogeneous distribution, and their lower contrasts (compared to starspots). Furthermore, the impact of other astrophysical noise sources (such as supergranulation, e.g. \citealt{Meunier2019_SG, Meunier2020_SG, Moulla2023, Lakeland2024}) requires further investigations. On top of these considerations, studies (e.g. \citealt{Hall2018}) have indicated that several years of RV observations are required in order to confirm, through RV mass-measurements, an Earth analogue (i.e. Earth-like planets in Earth-like orbits around solar-like stars). These long observational baselines also imply that the effects of long-term stellar activity cycles on RV measurements need to be better understood.

A number of techniques and observation strategies have been developed to identify and/or mitigate the effects of stellar variability on RV measurements (e.g. \citealt{Thompson2017, Dumusque2018, Meunier2019, Crass2021, Cameron2021, Cretignier2021, Cretignier2024, Larue2025}), and while great progress has been made in reducing the RV variability induced by stellar activity, none of these routes have yet successfully demonstrated the capability of recovering an Earth-analogue RV signal (e.g. see \citealt{Zhao2022}). These ongoing challenges highlight the need to deepen our understanding of how stellar activity impacts the observed spectra, alongside developing more robust stellar activity indicators that can precisely capture the sources of RV variability. 

For example, some studies have searched for activity-sensitive lines by examining correlations with established activity indicators (e.g. \citealt{Wise2018, Ning2019, Jiang2024, Gomes2025}). However, widely used diagnostics such as Ca II H \& K predominantly trace chromospheric variability, whereas the lines from which RVs are measured originate in the photosphere. Activity indicators that respond to photospheric changes are therefore likely to track the RV variations more accurately. These new photospheric indicators could then be used to better guide stellar activity mitigation techniques, such as multivariate Gaussian processes (GPs - e.g. see \citealt{Rajpaul2015, Aigrain2023, Klein2024}). In turn, these may help better understand the stellar physics driving such variations, allowing for an improved interpretability of RV modelling results as well as providing insights and potential new avenues for mitigating stellar activity.

This paper advances the work first presented by \cite{Thompson2017}, who applied spectral ratio analysis (SRA) to compare observations of $\alpha$~Cen~B taken during magnetically active and inactive periods with the High Accuracy Radial velocity Planet Searcher (HARPS) spectrograph. That study revealed a host of stellar (photospheric) lines that exhibited large variations correlated with \logRHK. The observed features in the ratio (active over inactive) spectra showed distinct morphologies, providing insight into the underlying stellar physics. Furthermore, several of these features also showed measurable velocity shifts, which simple modelling attributed to facular regions rotating across the stellar disc. This was later confirmed by \cite{Thompson2020} using HARPS-N solar telescope observations, whereby features in the disc-integrated ratio spectra of the Sun were directly tied to facular regions observed on the (resolved) solar surface.

Since the presence of faculae is an important astrophysical noise source that suppresses the convective blueshift and imprints velocities on the measured RVs, the direct link between faculae and the photospheric line-variability measurements uncovered by \cite{Thompson2017,Thompson2020} -- which includes velocity information -- merits further investigation. In this work, we apply SRA to a sample of \nstar~relatively inactive stars with long-term monitoring on the HARPS spectrograph that exhibit magnetic activity cycles. Our main objective is to test the utility of new activity indicators (derived from SRA features) for mitigating stellar activity induced RV variability. Section~\ref{sec:obs} describes our target selection, the SRA process, and outlines our method for identifying \feat\ from which we derive a new set of activity indicators. In Section~\ref{sec:SRAindicators}, we extract amplitude and velocity information contained in the SRA features across the full time-series of each star, and assess how these indicators vary with stellar activity by comparing them to the chromospheric \logRHK. Section~\ref{sec:RVmodel} quantitatively assesses the predictive power of the SRA indicators for explaining activity-induced RV variations. We compare their performance to the traditional \logRHK\ index and other cross-correlation function-based parameters such as the bisector span (BIS -- \citealt{Queloz2001}) and the full width at half-maximum (FWHM) using both a simple regression and a random forest, and evaluate their potential to improve stellar noise mitigation. Additionally, we perform planetary injection recovery tests to evaluate the potential of SRA to improve stellar noise mitigation and its limits. Finally, we discuss these results and present future prospects in Section~\ref{sec:conclusion}.

\section{Observations and Data Analysis}\label{sec:obs}
For this study, we have restricted our analysis to spectral types ranging from G to mid-K dwarfs. This spectral-type range mirrors that of upcoming RV surveys (such as the Terra Hunting Experiment\footnote{\url{https://www.terrahunting.org/goals.html}}, see e.g. \citealt{Baker2021}), which aim to identify potential Earth-like planets with relatively long orbital periods -- where the stellar variability problem is most acute. This range also covers the stellar parameter space where Earth-mass planets in the temperate zones of their host stars have largely gone undetected in previous surveys. Furthermore, following \cite{Costes2021}, we also concentrate on relatively inactive (likely faculae-dominated, e.g. see \citealt{Radick2018}) stars, as we expect these to represent the types of stars most suitable for detecting such planets.

\subsection{Target selection}\label{subsec:targets}
To perform the analysis presented in this paper, we used archival data from the HARPS spectrograph \citep{Mayor2003} mounted on the ESO 3.6 m telescope at La Silla Observatory, Chile. In terms of target selection, the criteria used were as followed:
\begin{itemize}
    \item Only HARPS data taken prior to BJD 2,457,150 were considered in order to avoid systematic offsets due to a fibre-upgrade intervention that required the HARPS vacuum enclosure to be opened. This intervention resulted in an offset between measurements taken before and after the upgrade. Following the approach of \cite{Costes2021}, the archival data were further vetted to remove outliers and spectra potentially affected by saturation. Since the analysis presented in this work involves detecting relatively small spectral variations (often lying at the $\sim$per-cent level or less) we also imposed a signal-to-noise ratio (SNR, taken from the 55$^{\mathrm{th}}$ order, i.e. $\sim$5800~\AA) lower threshold of 100 per individual observation in order to filter out lower-quality data.
    \item Our analysis makes use of spectra that are averaged on nightly and seasonal timescales (see Section~\ref{subsec:process}). To further guarantee adequate data quality in these combined spectra, we also set a requirement of minimum weighted SNRs for the nightly and seasonal averages. Based on the features identified for the Sun in \cite{Thompson2020}, we selected a minimum weighted SNR of 150 for the nightly averages. As we expect to average over one or more stellar rotational periods for the seasons, we selected a minimum weighted SNR of 500 for the seasonal averages.
    \item In order to study the long-term variations, the resulting baseline of observations must be long enough such that any underlying stellar activity cycle (or a decent portion of it) would be covered with reasonable sampling. Thus, we selected stars for which the observations that passed the previous cuts result in a baseline exceeding four years, and with observations taken during at least five different observing seasons.
    \item Another restriction was made to select relatively inactive (faculae-dominated) stars that nevertheless showed appreciable magnetic activity cycle variability. Since SRA relies on measuring the ratio between spectra taken at a period when the star was in a state of (relatively) high-activity in comparison to a spectrum taken in a low-activity state, such activity variations are required. Therefore, we only included stars with a median \logRHK\ $ < -4.80$ (as measured following the procedure outlined in \citealt{Lovis2011}) and activity cycles showing variations in \logRHK\ of $\Delta_\mathrm{act} > 0.05$, where $\mathrm{\Delta_{act}}$ is defined as the difference between the maximum and minimum seasonally binned \logRHK\ values (see details in Section~\ref{subsec:process}). 
    \item A final selection criterion was introduced to ensure sufficient nightly sampling for investigating shorter-term stellar variability (see Section~\ref{sec:RVmodel}). We retained only those stars with `high-activity' observations (i.e. above the low-activity template, see Section~\ref{subsec:process}) obtained on more than 50~separate nights, with those nights distributed over the different observing seasons. We note that this condition does not guarantee significant temporal coverage within a season, as the spectra may have been obtained on a single night. As a result, while our selection prioritises data quantity, it does not necessarily capture intra-seasonal variability such as stellar rotation -- a limitation imposed by the real-world sampling cadence of most RV surveys to date. 
\end{itemize}

Out of the 6165 stars probed from the public ESO archive\footnote{\url{http://archive.eso.org/wdb/wdb/adp/phase3 main/form}}, the above selection criteria yielded a final sample of \nstar~stars that are listed in Table~\ref{tab:1}. This table includes their spectral type (ranging from G2V to K6V), $B - V$ colour index (from SIMBAD\footnote{\url{https://simbad.u-strasbg.fr/simbad/}}), their $v \sin i$, total number of useful spectra, number of unique high-activity observing nights and seasons (see Section~\ref{subsec:process}), the median number of nights per season, and the total time-span of the observations. Table~\ref{tab:1} also reports the median \logRHK\ activity level and activity range ($\mathrm{\Delta_{act}}$). The seasonal and nightly averaged SNRs and the number of \feat\ detected for each star (described in Section~\ref{subsec:process}) are also presented in Table~\ref{tab:1}. Finally, all the reduced data used in this study were obtained from the public ESO archive and were processed with the HARPS Data Reduction Software (DRS) version~3.5.

\subsection{Data preparation and spectral ratio process}\label{subsec:process}
In this study we used the HARPS e2ds spectra -- a minimally processed, order-by-order DRS product -- to avoid repeated interpolation onto different wavelength grids later in the analysis. However, we note that the use of the 1D spectra also provided by the DRS (which are interpolated onto a fixed wavelength grid) yielded similar results.

The first step in the data preparation was to mask telluric lines in the spectra. We used \texttt{SkyCalc} to generate a telluric model for each star and for telluric lines deeper than 0.05\%, we masked out a 5~km~s$^{-1}$ region centred on each identified line. This conservative width was chosen to minimise the risk of telluric residuals contaminating the subsequent analysis, resulting in the removal of approximately 8\% of pixels over the full wavelength range.

The second step involved removing known RV shifts in order to align the observed spectra onto a common wavelength grid in the stellar rest frame -- a requirement for performing SRA. It is important to note that we did not use the RVs derived from the DRS cross-correlation functions (CCF), as they include apparent shifts due to activity-induced line-shape changes. Removing these shifts would then affect the activity-related signals we aim to isolate with SRA. Instead, we applied corrections for the barycentric radial velocity motion and secular acceleration of each star (see \citealt{Kamp1986, Choi2013}). We also removed RV signals due to known planets listed in planetary catalogues, including exoplanet.eu\footnote{exoplanet.eu: \url{http://exoplanet.eu/catalog/}} and the NASA exoplanet archive\footnote{NASA exoplanet archive: \url{https://exoplanetarchive.ipac.caltech.edu/}}. Following \cite{Costes2021}, and given that the time baselines of the stars in our sample often exceed those used to determine the published orbital solution, we re-modelled each planetary system using \texttt{allesfitter}\footnote{\url{https://github.com/MNGuenther/allesfitter}}. To mitigate long-term stellar activity variations, we included in our analysis linear correlations with several activity proxies. These indicators included the CCF BIS, the CCF FWHM, the CCF contrast, and the area of the Gaussian used to fit the CCF (see \citealt{Costes2021}). We then used these updated orbital fits to remove the RV signals of known planets. Ideally, this process would only leave RV variability associated with instrumental systematics and stellar activity. We note, however, that undetected and/or imperfectly removed planets may still introduce residual RV signals unrelated to stellar activity. As we demonstrate later in Section~\ref{sec:RVmodel}, small residual RV shifts have minimal impact on our SRA results. We also find that, while failing to remove known planetary signals does indeed degrade the correlations between the CCF RVs, \logRHK\ and the SRA parameters, it does not affect the main conclusions or results of this paper.

Once the known RV signals were removed, the third step was to reinterpolate each star's spectra onto a common wavelength grid using spline interpolation. This wavelength grid was defined separately for each star based on the spectrum with the highest SNR for that target. We tested alternative wavelength grids for each star and found no significant impact on the results. We note that the first five echelle orders were excluded due to their consistently low SNR, resulting in a final usable wavelength range of $\sim$~4025~\AA~to $\sim$~6800~\AA.

Where multiple spectra were available for a given night, we combined them using a weighted average, with weights equal to the square root of the flux. The resulting SNR of these combined spectra was calculated by accounting for the limiting noise floor of the daily master flat fields used in the reduction pipeline, which exhibit an average SNR $\sim 650$ (see e.g. \citealt{Mayor2003}). The measured nightly average SNR are reported in Table~\ref{tab:1}. We also applied the same weighted averaging procedure across all spectra obtained within a single observing season in order to boost the overall SNR and average over shorter-term variability associated with stellar rotation, and we refer to these combined spectra as `nightly' or `seasonal' spectra, respectively. For each nightly and seasonal spectrum, we applied the same weighted methodology to measure its respective \logRHK\ values.

\begin{figure*}
    \centering
    \includegraphics[width=1\textwidth]{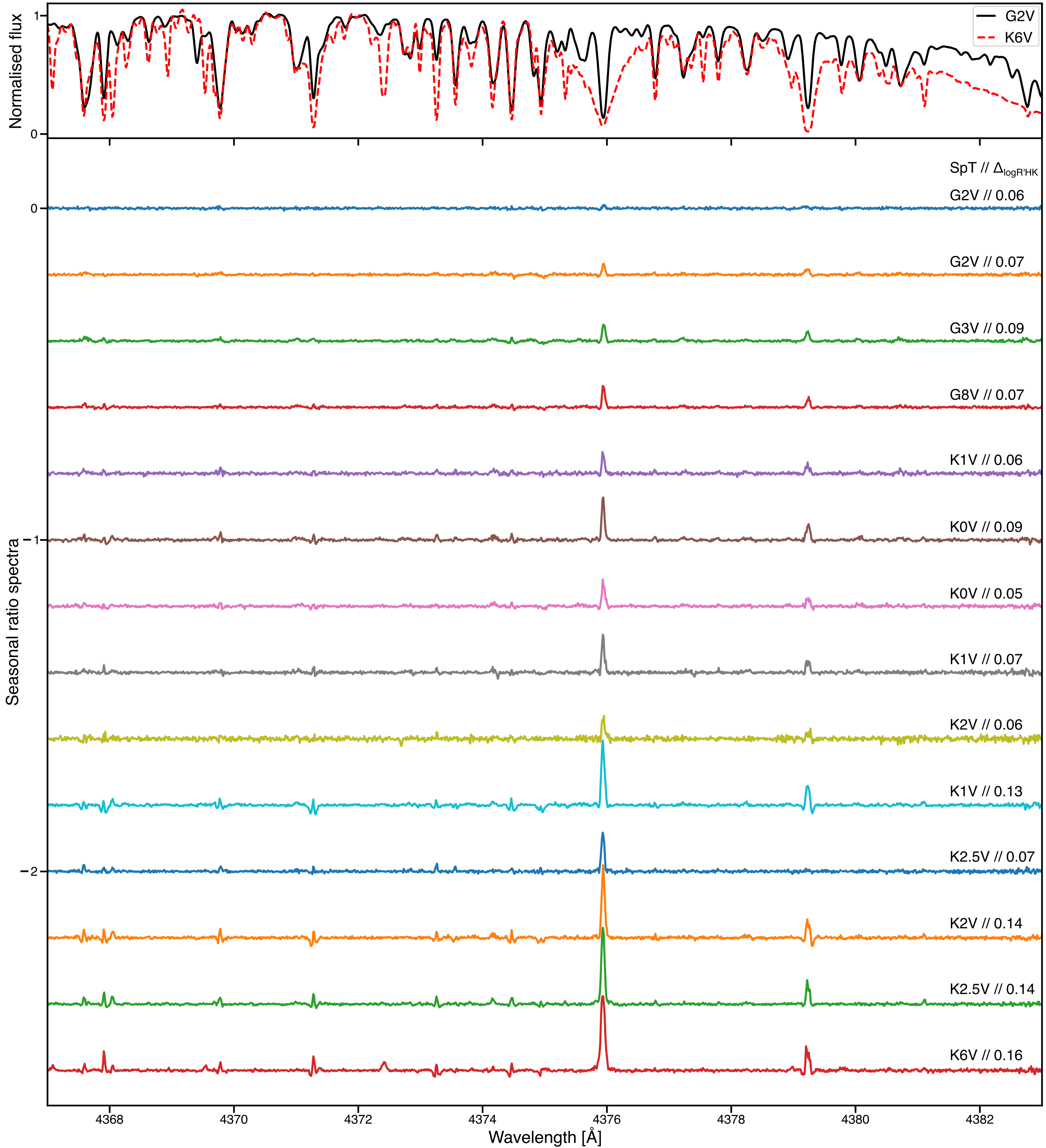}
    \caption{\label{fig:StarsFlux} A selection of seasonal ratio spectra for the \nstar~stars used in this study, ordered by $B - V$ (as presented in Table~\ref{tab:1}) and removed from the ripple-like structure. The ratio spectra were generated by dividing high-activity spectra by a low-activity template for each star (SRA). For comparison, all the spectra are shown in their rest wavelength and, in order to increase the signal-to-noise, we have combined the three most active seasonal ratio spectra for each star. The wavelength range has been restricted to between 4367.5~\r{A} and 4382.5~\r{A} in order to show some detail of prominent features. Two representative normalised spectra, from a G2- and a K6-type star, covering the same wavelength range as the ratio spectra are shown in the upper panel. The most prominent feature located around 4375.9~\r{A} corresponds to an extremely activity-sensitive Fe I line. The values to the right of each ratio spectrum indicate the spectral type of the target and the difference of activity as measured by \logRHK\ between the seasonal average spectrum and the low activity template ($\Delta_{\mathrm{logR_{HK}}}$), respectively. Each ratio spectrum has been offset vertically for clarity.}
\end{figure*}

Finally, each order of these nightly or seasonal spectra was normalised to one using the peak of the flux. Then, nightly and seasonal ratio spectra were created by dividing these spectra by a low-activity template spectrum. This template was selected for each star from the seasonally averaged spectrum with the lowest activity level. Since the e2ds spectra used were not blaze-corrected, the resulting ratio spectra exhibited some broad-scale, low-amplitude instrumental variations across the wavelength domain. Because of the low-frequency nature of these trends, comparative tests confirmed that the specific choice of correction, whether via low-order polynomial fitting or Gaussian filtering, did not significantly influence the results. We adopted the latter approach for this analysis, applying a Gaussian filter with $\sigma = 15$~pixels, and divided the ratio spectra by this smooth fit to flatten the remaining residuals.

As mentioned above, the SRA methodology relies on dividing high-activity spectra by a low-activity template. Thus, we decided to remove any nightly spectra with a \logRHK\ lower than that of the low-activity template of each star from this study. While this conservative approach was adopted to avoid introducing features with potentially different (e.g. inverted) morphologies than expected, it resulted in an average loss of $\sim$20\% of observations per star. The impact of including cases where spectra have activity levels lower than that of the inactive template will be explored in a future study.

Figure~\ref{fig:StarsFlux} presents a subset of the seasonal ratio spectra generated by SRA, ordered by $B - V$ (as presented in Table~\ref{tab:1}) after removal of additional systematics (see Section~\ref{subsec:ripple}). Each spectrum represents the weighted average of the three most active seasonal ratio spectra for that star (hereafter denoted SRA$_{\mathrm{Template}}$), which helps increase the SNR and emphasise the activity-related features. To aid direct visual comparison between the spectra, the systemic velocities of the stars have been removed and we only show wavelengths between 4367~\r{A} and 4383~\r{A} centred around the strong Fe~I feature at 4375.9~\r{A}. Two representative normalised spectra, from a G2- and a K6-type star, covering the same wavelength range are shown in the upper panel. The spectral type of each star and the difference in activity as measured by \logRHK\ between the seasonal average spectrum and the low activity template are also indicated in Figure~\ref{fig:StarsFlux}. As can be seen, the ratio spectra reveal the spectral imprint of stellar activity, which principally manifests itself as apparent emission peaks. Visual inspection suggests that the strength of the different features varies with both the spectral type and the amplitude of the star's magnetic cycle. However, we note that the morphologies of these features are more complex than this, as discussed by \cite{Thompson2017, Thompson2020}. We also considered whether observational parameters -- such as the number of spectra, total observational baseline, or median SNR -- influenced the apparent strength of activity features. However, no significant correlations were found.

All the seasonal ratio spectra shown in Figure~\ref{fig:StarsFlux} (but covering the full wavelength range) are available in electronic form via an online atlas hosted at Zenodo \citep{Costes2026_dataset}. For each star, the nightly and seasonal SRA spectra per order, with their respective errors, wavelength, BJD and \logRHK, are also available in that atlas. Additionally, all of the SRA detected features (see Section~\ref{subsec:atlas}) are also provided, with their wavelength, amplitude, width and potentially associated line obtained from the VALD line list. This resource may aid the identification of additional activity-sensitive features, support line-by-line analyses, or help benchmark magnetohydrodynamic simulations of active stellar photospheres.

\subsection{Ripple structures}\label{subsec:ripple}
Upon inspecting the resulting ratio spectra (SRA residuals), we identified systematic `ripple'-like structures present across all the stars in this study. These `ripples' have typical amplitudes of $\sim$0.05 to 1\% and wavelength scales varying between $\sim$0.5 to 2~\AA\ (see below). The most prominent of these `ripples' originate from a filter that was inserted  into the light path after the Tungsten lamp within the HARPS calibration unit in 2007. The introduction of this filter resulted in Fabry-Perot-like interference that affected the flat-field calibrations (see \citealt{Cretignier2021}), but was thought to have been resolved in 2009 after an upgrade to HARPS. However, in this work we still detect systematic `ripple'-like structure in data taken after 2009 when the upgrade was completed, although they are smaller in amplitude. Hence, the origin of these `ripple'-like structures remains unclear as they appear irregularly, on most nights but not all, and affect most spectral orders, though their exact form varies between observations. Using different low-activity templates to create the ratio spectra did not eliminate these ripples. Similar artefacts have also been found for other instruments. For example, `wiggle' patterns have been observed in ESPRESSO spectra with fixed wavelength scales of 1~\AA\ and 30~\AA, which are likely caused by an interference pattern induced by the coud\'{e} train optics (see e.g. \citealt{Tabernero2021, Bourrier2025}). \cite{Thompson2020} reported similar ripple patterns in ratio spectra created using the HARPS-N solar telescope, and attributed them to etaloning from optical elements. Similarly to \cite{Thompson2020}, we confirm that these ripples were not introduced by our wavelength interpolation method.

Since these ripples introduce structured noise into the ratio spectra -- and our analysis seeks to identify subtle line-profile variations due to stellar activity (see Section~\ref{subsec:atlas}) -- we developed a correction procedure to mitigate their impact. For each nightly and seasonal ratio spectrum created, we used the Generalised Lomb-Scargle (GLS) periodogram \citep{zechmeister2009generalised} to measure the frequency of the ripples. We found that the ripple pattern varied both with wavelength and also as a function of order in the sense that the ripple pattern was slightly different between orders where there was wavelength overlap. In general, the ripple pattern's frequency decreased towards longer wavelengths.

Unlike \cite{Thompson2020}, we did not use multiple sine waves to remove the ripple pattern as we found that this still left considerable residuals. Instead, we found that the ripples were better removed by fitting a single sinusoidal with a wavelength-dependent frequency. We adopted the following prescription for the ripple model:
\begin{ceqn}
\begin{equation}\label{eq:ModelRipple}
   M_\mathrm{ripple}(\lambda) = 
   A \times \sin\left(\frac{2\pi}{\Delta_{\lambda}} \times (\lambda - \lambda_{0,n}) \times \frac{\lambda_n}{\lambda}\right) + K~,
\end{equation}
\end{ceqn}
where $M\mathrm{_{ripple}}\left(\lambda \right)$ is the model of the ripple at a given wavelength $\lambda$, $\lambda_n$ is the central wavelength of order $n$, and $A$, $\Delta_{\lambda}$, $\lambda_{0,n}$ and $K$ are the amplitude, best-fit wavelength scale (i.e. the local periodicity), reference wavelength, and offset derived from the GLS periodogram, respectively. Additionally, we investigated these ripples in the context of thin-film interference theory (see e.g. \citealt{Born1999}). Specifically, for the case where the interference occurs in a beam perpendicular to the film, we used the following model: 
\begin{ceqn}
\begin{equation}\label{eq:thinfilm}
   M_\mathrm{tf}(\lambda) = 
   A \times \sin\left(2\pi \times \frac{2NL}{\lambda}\right) + K~,
\end{equation}
\end{ceqn}
where $M\mathrm{_{tf}}\left(\lambda \right)$ is the thin-film model of the ripple at a given wavelength $\lambda$, $N$ is the refractive index of the film and $L$ is the thickness of the film.
By fitting the observed `ripples' in the SRA residuals with this thin-film interference model, we were able to measure a consistent optical path length $NL = 1.5 \times 10^{7}$~\AA\ across the different orders for all our targets, where $N$ corresponds to the refractive index and $L$ is the thickness of the film. Assuming $N \sim 1.5$ for standard optical glass, this results in a component thickness of $L \sim 1$~mm. We note that this derived thickness of 1~mm is a typical size for filters that are used to attenuate the signal and rebalance the blue/red spectrum. While we are not able to fully understand the origin of these `ripples', we note that in terms of the resulting noise level, we find no significant differences after correcting the ripple pattern using the thin-film interference model in Equation~\ref{eq:thinfilm} or using Equation~\ref{eq:ModelRipple}.

\begin{figure}
    \centering
    \includegraphics[width=1\columnwidth]{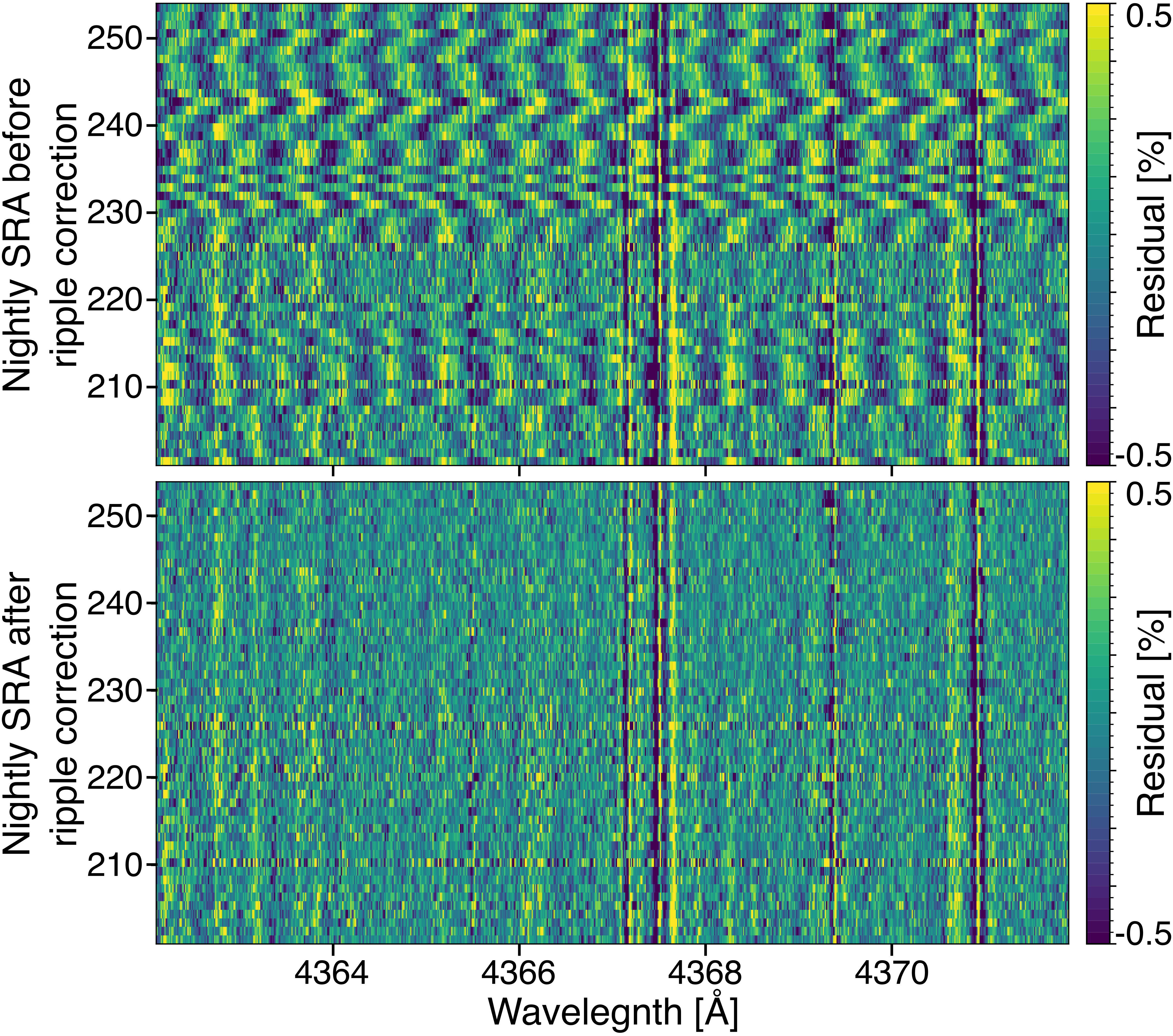}
    \caption{\label{fig:Ripple} Observation and removal of the `ripple'-like structure in SRA. The top panel presents a portion of the nightly ratio spectra for the target $\alpha$~Cen~B. As can be seen, some patterns appear irregularly on some of the nightly ratio spectra. The bottom panel presents the same portion of the nightly ratio spectra after the correction of the ripples using Equation~\ref{eq:ModelRipple}, enhancing the visibility of the activity-induced features, which appear as vertical lines.}
\end{figure}

The top panel of Figure~\ref{fig:Ripple} presents a portion of the nightly ratio spectra for the target $\alpha$~Cen~B. As can be seen, irregular pattern can be observed for some of the ratio spectra. The bottom panel of Figure~\ref{fig:Ripple} presents the same ratio spectra after removing the `ripple'-like structures using the equation above. Applying this correction reduced the standard deviation of the ratio spectra by a factor of $\sim$2. This enhanced the visibility of activity-induced features and enabled the identification of other weaker features that were previously obscured by this instrumental artefact. 

\subsection{Identification of activity-sensitive line variability}\label{subsec:atlas}
\begin{figure}
    \centering
    \includegraphics[width=0.99\columnwidth]{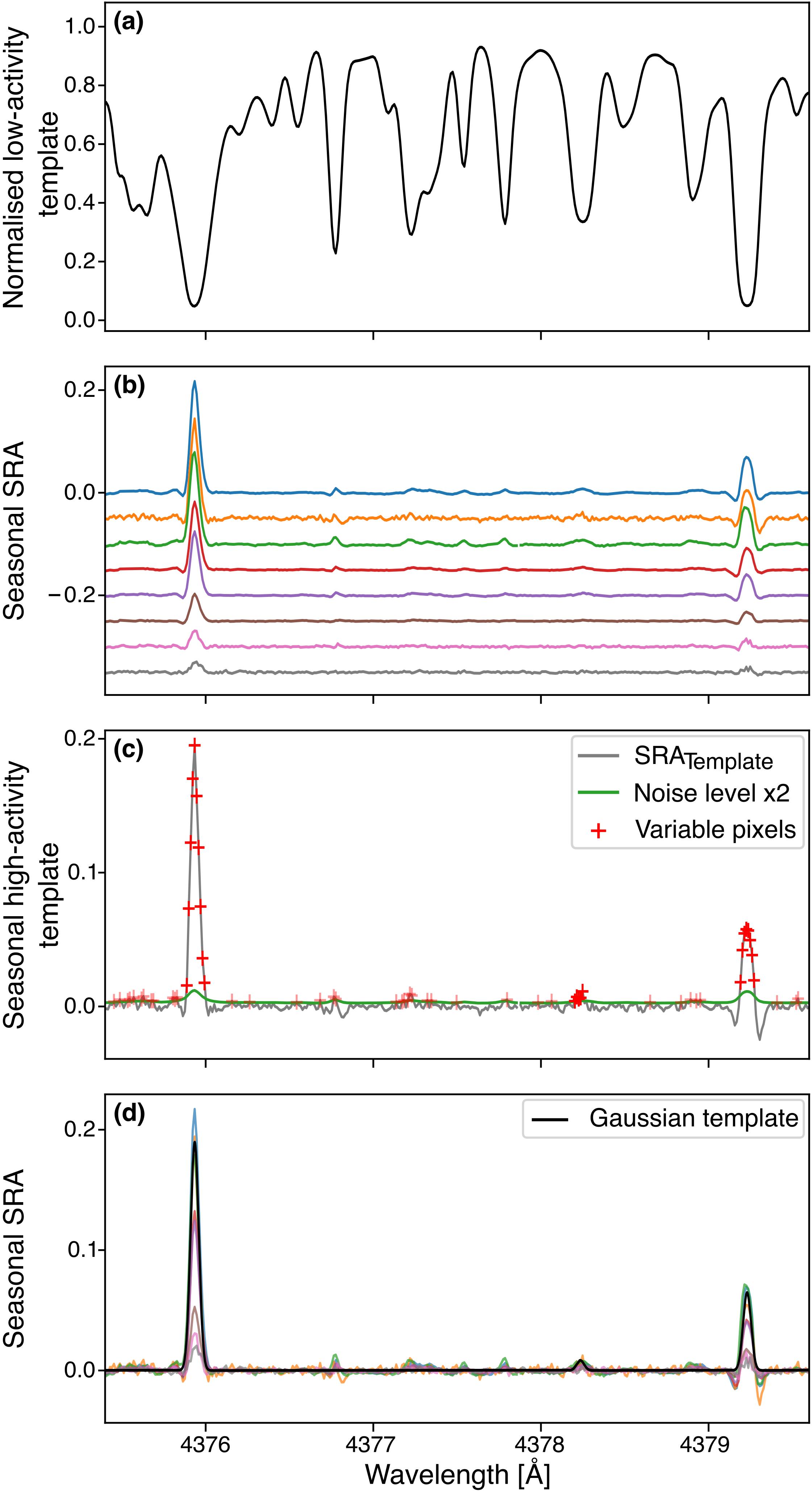}
    \caption{\label{fig:Feats} A graphical outline of the process of detecting \feat\ in the ratio spectra of $\alpha$~Cen~B. The first panel (a) presents the low-activity template used to create the nightly and seasonal ratio spectra. The second panel (b) shows the eight seasonal ratio spectra created, offset from one another for clarity and ordered by their median \logRHK, with the most active at the top. The third panel (c) represents the high-activity SRA$_{\mathrm{Template}}$ in grey. Pixels that have a value twice above the weighted noise level (shown in green) are highlighted by a red cross. Transparency was used to differentiate variable pixels that were not considered as a `true' \feat. The last panel (d) shows, in black, the Gaussian fits to the three \feat\ that were identified through this selection process, on top of the eight seasonal ratio spectra.}
\end{figure}

As part of our analysis we developed an automated method to identify spectral features that vary across our stellar sample. Feature identification was based on the seasonal ratio spectra, which have enhanced SNRs and therefore enable better identification of activity induced signals. Figure~\ref{fig:Feats} illustrates (for the K1V star HD128621, also known as $\alpha$~Cen~B) the process of identifying variable spectral features in the seasonal ratio spectra focusing on the wavelength region around the activity sensitive Fe I line at 4375.9~\r{A}. First, we used the low-activity template (i.e. the lowest-activity seasonally averaged spectrum, see panel (a) of Figure~\ref{fig:Feats}) to create both nightly and seasonal ratio spectra as described in Section~\ref{subsec:process}. Panel (b) presents the eight seasonal ratio spectra created for this target, ordered by their median \logRHK~activity level. As can be seen, features are clearly visible in the resulting ratio spectra, with the strongest feature (corresponding to variations of the Fe I 4375.9~\r{A} line) visually appearing to correlate well with \logRHK. 

Next, we used the high-activity SRA$_{\mathrm{Template}}$ and its associated weighted noise to identify pixels that show significant amplitude variability. Thus, any pixel that has a value twice that of the noise level was flagged (see panel (c) in Figure~\ref{fig:Feats}). These variable pixel candidates were then analysed for clustering to identify coherent features. A sliding window of four continuous pixels was used to identify feature candidates, based on the assumptions that such features would cause multiple nearby pixels to exceed a set standard deviation threshold. Within each window, if two or more pixels exceeded the threshold, the wavelength range considered to correspond to a feature was extended. The window was then advanced to the last identified variable pixel, and the process was repeated. The initial bounds of the feature were defined as the range from the first to the last pixel that satisfied this clustering criterion. Isolated variable pixels were discarded as likely noise. Additionally, we required each cluster to contain at least two consecutive variable pixels above zero in the SRA$_{\mathrm{Template}}$ for the feature to be considered as a valid candidate. While this method for identifying features is not perfect, we found that it was sufficient for reliable detection. Since our goal in this proof-of-concept is to isolate and analyse unambiguous activity-sensitive features, rather than fit their exact profiles, the emphasis remained on robust peak identification.

Finally, a Gaussian was fitted to each of the identified features in the SRA$_{\mathrm{Template}}$ (shown as black curves in the last panel (d) of Figure~\ref{fig:Feats}). To avoid biases and fitting issues, a final cut was applied by discarding any feature with a FWHM below 2.5~km~s$^{-1}$, which is comparable to the instrumental resolution of HARPS. The remaining features were considered as `true' activity-sensitive features. As an example of our identification process, three distinct \feat\ of different amplitudes that passed this selection process can be seen in Figure~\ref{fig:Feats} at 4375.9~\AA, 4378.3~\AA, and 4379.2~\AA, while the other signals showing variable pixels were rejected as outliers.

The velocity shift and amplitude of each of these identified features were measured individually in the nightly and seasonal ratio spectra using Gaussian fitting. This feature-by-feature approach was taken as the amplitudes of different features appear to vary out of phase with each other (these variations among different features will be explored in a future paper).

Since these features trace active regions on the stellar surface (analogous to the apparent `emission' bumps used to map the spatial locations of starspots in Doppler imaging, see \citealt{Thompson2020}), their velocities are expected to vary due to the different rotational phases that the active regions are observed at. Physically, the maximum extent of these velocity variations must be less than the magnitude of the projected stellar equatorial rotation velocity ($\pm v \sin i$). Since all the stars in our sample are slow rotators and hence have low $v \sin i$'s (that are difficult to measure accurately) we adopted a conservative velocity shift cutoff of $\pm4.5$~km~s$^{-1}$.

Using this identification process, hundreds of features were detected for each star (as presented in the last column of Table~\ref{tab:1}). This highlights the potential of SRA to reveal rich information on the impact of stellar activity on the spectra of even relatively quiet stars. Fully exploiting and interpreting this will require further investigation beyond this paper. In particular, the diverse range of morphologies and temporal evolution of these features warrant deeper study (see for instance Hobbs et al. in prep, and Yaptangco et al. in prep).

We note that no specific lines were excluded besides those that fell outside the wavelength range used (4025 -- 6800~\AA, such as the Ca~\textsc{ii} H \& K lines) and those affected by telluric removal (e.g. H$\alpha$ and the Sodium D-doublet). Hence, if the remaining broad lines provided well-defined features (i.e. passing the selection criteria), they were included (e.g. as was the case for the Mg~\textsc{i} b$_{1}$ \& b$_{2}$ lines). Conversely, the H$\beta$ and H$\gamma$ lines did not pass the selection criteria. It is also worth noting that, due to the high number of detected features for each star, the contributions of individual features are strongly diluted. Therefore, the inclusion of a handful of lines with substantially different feature morphologies does not appreciably affect the results presented in this work. In the rest of the paper, we will focus our study on the amplitudes and velocity shifts of these features, and how these vary across both short and long time scales.

\section{Photospheric response variations}\label{sec:SRAindicators}
As previously shown by \cite{Thompson2017, Thompson2020}, the amplitudes of many of the features identified via SRA in both $\alpha$~Cen~B and the Sun were strongly correlated with \logRHK. Building on this, we investigated whether similar correlations hold across the range of spectral types in our stellar sample, while also examining the behaviour of the associated feature velocity shifts.

Using the nightly and seasonal Gaussian fit amplitudes and shifts measured for all the features for each star in our sample, we computed two summary metrics designed to capture broad trends in the photospheric-line variability revealed by SRA. For each night and season, we calculated the weighted average of the Gaussian fit amplitudes and shifts using their respective fitting errors. This yielded two `global' SRA variability indicators, per star, that we refer to throughout the rest of this work as the SRA$_{\mathrm{Feature}}$ amplitude and SRA$_{\mathrm{Feature}}$ shift. These metrics were then compared with the corresponding \logRHK\ variations for each star. 

We emphasise that this result is a grossly simplified view of the photospheric-line variability captured by the SRA process. Collapsing the wide variety of line-responses to stellar activity into a single pair of metrics inevitably discards much of the rich, feature-specific information available. Nonetheless, we adopt this simplified approach here as a practical first step because the full complexity of the variability is not yet fully understood. A deeper exploration of the feature-by-feature behaviour and their physical interpretation is deferred to future works.

\begin{figure*}
    \centering
    \includegraphics[width=\textwidth]{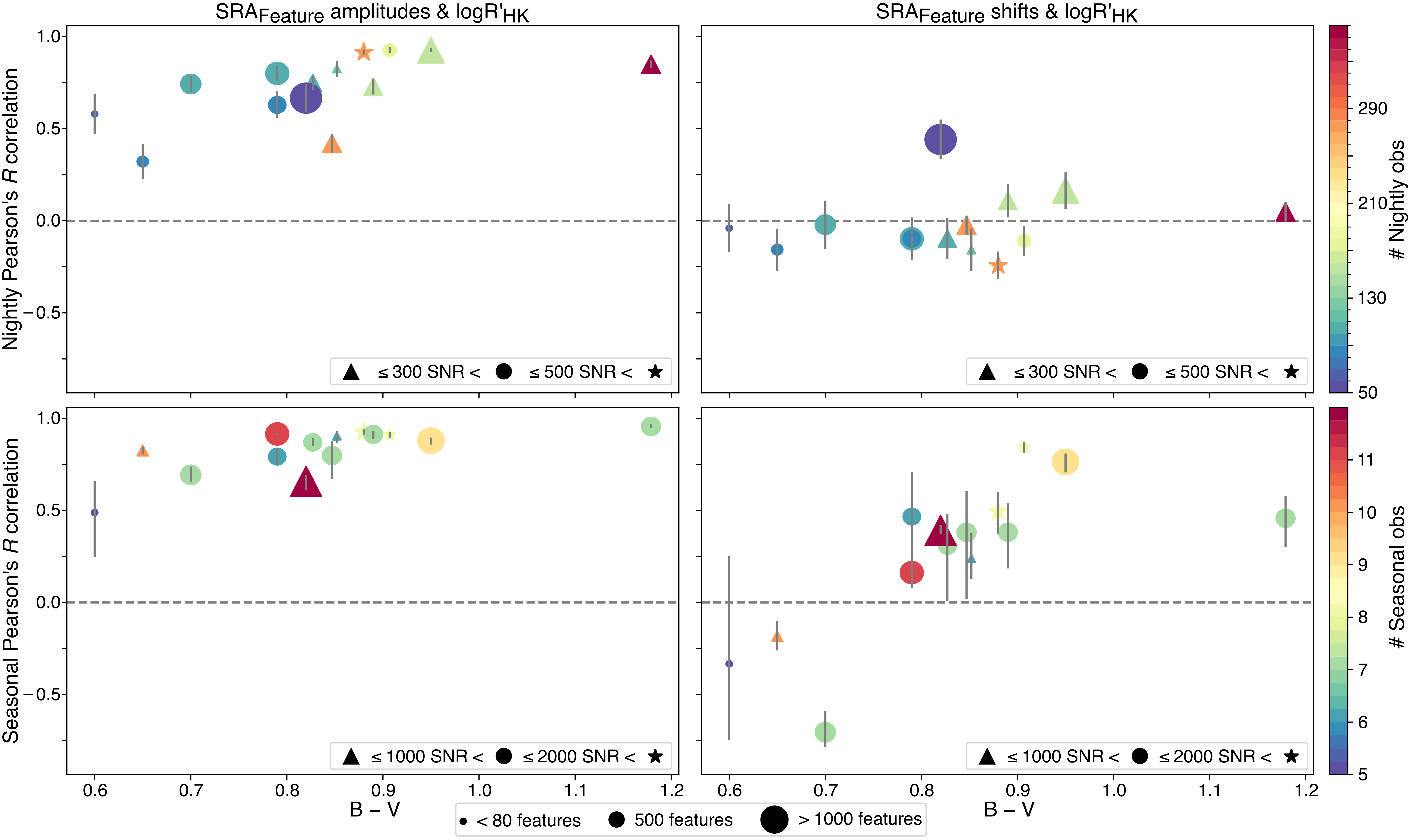}
    \caption{\label{fig:corrlog} Pearson's~$R$ correlation between the SRA parameters, SRA$_{\mathrm{Feature}}$ amplitudes and shifts, and \logRHK\ for each star. The upper panels present the correlation over the short time scale using the nightly data. Conversely, the lower panels present the correlation over the long time scale using the seasonal data. A different shape was assigned to represent the median value of the weighted average SNR in the 55$^{\mathrm{th}}$ order for the nightly and seasonal panels. Additionally, the size of each point represents the number of \feat\ used in the analysis, while the colour of each star represents the number of nightly and seasonal observations per star. Vertical grey lines represent the variability of the measured correlations. These were used in order to identify possible cases of bias.}
\end{figure*}

The left panels of Figure~\ref{fig:corrlog} present the Pearson's rank correlation coefficient, $R$, between the SRA$_{\mathrm{Feature}}$ amplitudes and the corresponding \logRHK values for the \nstar~stars in our sample, while the right panels show the corresponding values for the SRA$_{\mathrm{Feature}}$ velocity shifts. The correlations were computed for both the nightly (top panels) and seasonal (bottom panels) data for each star. We used the Pearson's $R$ correlation coefficient, which assesses linear relationships between two quantities, as a crude tool to determine whether any correlation is present, and if so, whether it is strong or weak. However, we should note that we have performed the same analysis using the Spearman correlation (which quantifies monotonic relationships) and similar results were found. The shape of each marker indicates the median SNR of the nightly and seasonal spectra in the 55$^{\mathrm{th}}$ order. We remind the reader that several spectra during a single night and season may be combined in a weighted mean, leading to reported SNRs that would be above the saturation limit of a single HARPS exposure. Additionally, the marker size reflects the number of \feat\ used in the analysis, while the colour of the markers represents the number of nightly and seasonal observations per star. As such, stars represented by a smaller, blue, triangle marker have a lower number of \feat, have been observed for fewer nights/seasons, and have a lower SNR, and therefore the results for these stars should be treated with more caution. Finally, uncertainties on these correlations are shown by the vertical grey lines on each marker. Due to the difference in the amount of data used between the nightly and the seasonal dataset, these were computed in a different manner in each case. For the nightly data, the correlation was measured one hundred times after randomly removing 10\% of the data, and the standard deviation that was computed from these results was adopted as the uncertainty on the correlation. For the seasonal dataset, the uncertainties represent the 0.95 confidence level of the jackknife estimate. In the following subsections we discuss the correlations between \logRHK\ and the two SRA metrics in turn.

\subsection{Nightly and seasonal correlations between the feature amplitudes and \logRHK}\label{sec:famp_variations}
For the nightly data, a strong positive correlation between the SRA$_{\mathrm{Feature}}$ amplitude and \logRHK\ is evident for the majority of stars (top-left panel Figure~\ref{fig:corrlog}). This is consistent with expectations. As stellar activity increases, this typically results in the presence of more and/or larger active regions, enhancing the distortions in the spectral lines and leading to greater differences between the observations and the low-activity template. As a result, higher activity levels correspond to stronger features in the ratio spectra.

Interestingly, the strength of the correlation appears to weaken for earlier type stars. While this trend could be somewhat influenced by the sparser data coverage for the two hottest stars in our sample, it may hint at an underlying astrophysical effect. One possibility is that the stronger correlations observed in the later-type stars arise from a tighter coupling between the chromosphere and photosphere due to the more compressed vertical extent of their atmospheres. This, in turn, could lead to a stronger coherence between the amplitudes of the (SRA-derived) photospheric and the chromospheric (\logRHK) indicators in later-type stars. If this interpretation is correct, it may have important ramifications for astrophysical noise mitigation techniques that rely on \logRHK\ as a proxy for activity-induced RV signals. In particular, for earlier type stars (where the chromosphere and photosphere responses to stellar activity appear to be more decoupled) the reliability of \logRHK-based corrections may be diminished.

For the seasonal data (bottom left panel of Figure~\ref{fig:corrlog}), most stars in our sample show a strong positive correlation (typically $R >$ 0.7) with the change in \logRHK\ over the magnetic cycle. The weaker correlation observed among the earlier-type stars in the nightly data is also seen in the seasonal results, but we urge caution due to the sparse dataset and the low amount of data point used in the seasonal correlation (as low as 5, see the selection criterion in Section~\ref{subsec:targets}). Regardless, Figure~\ref{fig:corrlog} clearly shows that the SRA$_{\mathrm{Feature}}$ amplitudes strongly track long-term stellar activity variations associated with the magnetic cycle.

\subsection{Nightly and seasonal correlations between the features velocity shifts and \logRHK}
For the nightly data, no strong correlation is observed between the SRA$_{\mathrm{Feature}}$ velocity shifts and the variations in \logRHK\,(top right panel Figure~\ref{fig:corrlog}). This lack of correlation is unsurprising, since we believe that the SRA$_{\mathrm{Feature}}$ velocity shifts are mostly tracking the apparent motion of active regions as they rotate across the visible stellar hemisphere. As a result, the largest absolute magnitude of the SRA$_{\mathrm{Feature}}$ velocity shifts (with respect to the systemic velocity) should occur when the active region is located at the stellar limbs (e.g. with a maximum velocity shift of $\pm v\sin i$), but where the overall activity level (as captured by \logRHK) is comparatively low due to fore-shortening. Conversely, the lowest SRA$_{\mathrm{Feature}}$ shift should occur when the active region is near the centre of the disc when the activity level would peak. 

Furthermore, the measured SRA$_{\mathrm{Feature}}$ velocity shift depends mainly on the position of the active regions, not their size or contrast. A small active region at disc-centre (for example) would have a similar velocity to a much larger one at the same spatial location, but \logRHK\ would differ between the small- and large-active region cases. This combination of geometric effects and the degeneracy between the active region's size/contrast and its projected rotational velocity should act to greatly weaken any correlation between the SRA$_{\mathrm{Feature}}$ velocities and \logRHK. The presence of a complex distribution of faculae across the stellar surface would further degrade any correlation.

In contrast to the nightly results, the seasonally averaged data show moderate correlations between the SRA$_{\mathrm{Feature}}$ velocity shifts and \logRHK\ for the majority of the stars (bottom right panel Figure~\ref{fig:corrlog}). Additionally, we note a potential trend between the hottest stars being anti-correlated and the cooler stars being positively correlated. However, after close inspection, the G-type star with a Pearson's $R$ correlation of $\sim -0.7$ has two seasonal outliers that are strongly driving the anti-correlation. Therefore, due to potential additional outliers, the sparse amount of data and large uncertainties, we disregard the three hottest stars in this section, focusing only on the bulk of the stars presenting a moderate positive correlation.

As outlined above, this moderate positive correlation between the SRA$_{\mathrm{Feature}}$ velocity shifts and \logRHK\ is rather surprising as we do not expect a relationship between the two. The primary difference between the seasonal and nightly ratio spectra is that, in the former, the shorter-term rotational modulation signals will be (to some extent) averaged out. This may then allow more slowly varying signals occurring over the stellar activity cycles to become apparent.

If these stronger positive correlations are astrophysical in origin, then one possible explanation could be a long-term variation in the magnitude of the convective blueshift suppression for the cooler stars. Over the course of a magnetic cycle, increases in surface magnetic field strength within active regions would further inhibit the convective motions, reducing the upflow velocities compared to epochs of weaker field strengths. Magneto-hydrodynamical simulations spanning a range of spectral-types \citep{Beeck2015a} indeed show a modest effect where upflow velocities decline as the field strength increases. Since the SRA$_{\mathrm{Feature}}$ velocity shifts probe the velocities underlying the active regions, any systematic change in the underlying convective flow velocities should also leave an imprint in these measurements. In this case, we should see the measured SRA$_{\mathrm{Feature}}$ velocities progressively redshift (to more positive velocities) as stellar activity increases, leading to a positive correlation with \logRHK -- as is seen in Figure~\ref{fig:corrlog}. While speculative and requiring more detailed investigation, this could be important as the suppression of convective blueshift has been shown to be one of the dominant astrophysical noise contributions in the solar case with regards to the detectability of Earth-analogue planets (e.g. \citealt{Meunier2010, Meunier2024}). Knowing how the suppression of convective blueshift may vary over an activity cycle could be valuable. The link between the \feat\ properties and the potentially variable magnitude of the suppression of the convective blueshift over stellar activity cycles will be studied in a follow-up paper.

\section{Astrophysical noise mitigation via SRA}\label{sec:RVmodel}

The main motivation of this paper was to compare the performance of SRA-derived activity indicators against other widely used proxies, such as the chromospheric activity index \logRHK, or the CCF BIS and FWHM indicators, for mitigating the effects of stellar activity in RV measurements. For simplicity, we adopted \logRHK\ as our benchmark in this study, over other indicators, as \logRHK\ is considered as a standard proxy for tracing stellar activity in RV time series (e.g. \citealt{Lienhard2023, Cretignier2024}). For completeness, the results obtained for the other activity indicators are discussed in Section~\ref{subsec:alphacenb}~\&~\ref{subsec:allstars}.

While the utility of \logRHK to mitigate the effects of stellar activity is well-proven, it measures chromospheric emission, which does not necessarily provide a clean representation of the variability in the photospheric lines. By contrast, since SRA extracts activity signals directly from photospheric lines (where the RV measurements are also derived), SRA-derived activity indicators should offer a more direct link to the sources of RV perturbations. In addition, it provides both amplitude and velocity information. In this sense, \logRHK\ can be thought of as a scalar activity indicator indicating the strength of chromospheric activity, whereas SRA-derived activity indicators may act more like a vector quantity probing photospheric processes -- encoding both a magnitude (via SRA$_{\mathrm{Feature}}$ amplitude) and a direction (in velocity space via the SRA$_{\mathrm{Feature}}$ shift). In principle, combining these vector-like measures derived from photospheric lines should capture a broader range of activity-driven RV signals than a scalar index like \logRHK\ that stems from processes occurring in the chromosphere.

In this Section, we assess these SRA-derived indicators in their ability to mitigate astrophysical noise in RVs, benchmarked against \logRHK. It is worth reiterating that we have distilled the rich information from the SRA into just two summary metrics (SRA$_{\mathrm{Feature}}$ velocity shift and amplitude) in this proof-of-concept. This deliberate simplification discards the feature-by-feature diversity in line-shape responses to activity, and so the results presented here should be regarded as a conservative estimate of SRA’s potential capability for astrophysical noise mitigation. We first focus our analysis on $\alpha$~Cen~B as a case study using a linear regression model to compare the results of \logRHK\ with the SRA indicators. Then, we apply the same method to all \nstar~stars in our sample and measure the root mean square (RMS) of the RV residuals of these fits. We additionally confirm our results with another approach, using random forest regression to determine the importances of the different indicators for each star. Finally, we perform planetary injection recovery tests to evaluate the potential of SRA to improve stellar noise mitigation and its limits.

\subsection{A case study: Comparing SRA-derived activity indicators versus \logRHK\ via linear RV modelling}\label{subsec:alphacenb}
Using the SRA-derived features, we investigated how well the information from these measures tracks RV variations induced by stellar activity compared to \logRHK. In this section we assessed this by testing two multiple linear regression models -- one using \logRHK\ and the other using SRA$_{\mathrm{Feature}}$ amplitudes and velocity shifts -- assuming the following relationships:
\begin{equation}\label{eq:RVmodel}
\begin{split}
    &\mathrm{RV}_{\mathrm{\mathrm{log}\,R'_\mathrm{HK}}} = \alpha_1 + \alpha_2 \times \mathrm{log}\,R'_\mathrm{HK} + \alpha_3 \times \left(\mathrm{log}\,R'_\mathrm{HK}\right)^2\\
    &\mathrm{RV_{SRA}} = \beta_1 + \beta_2 \times \mathrm{SRA_{F.\ amplitude}} + \beta_3 \times \mathrm{SRA_{F.\ shift}}~.
\end{split}
\end{equation}
Including the \logRHK\ squared term ensured both models used the same number of free parameters for a fair comparison and also accounts for the non-linear relationship between \logRHK and the CCF RVs. These were then fit to the RV variations measured by the CCFs (as output by the DRS pipeline) after removal of any known planetary signals (see Section~\ref{subsec:process}).

We focused our analysis on the short-term variations by fitting the nightly data, analysing each season separately. This strategy avoids the complication that the relationship between the activity indicators and the stellar RVs can evolve over longer timescales due to changes in the active region properties or spatial distributions over the magnetic cycle (which is known to produce a hysteresis effect between RVs and activity indicators, see \citealt{Meunier2019}). The use of an additional second order polynomial in time was also tested ($+ \gamma t + \delta t^{2}$, following \citealt{Dumusque2017}), but no major differences were noticed as the time terms had negligible impact when modelling individual seasons.

\begin{figure*}
    \centering
    \includegraphics[width=\textwidth]{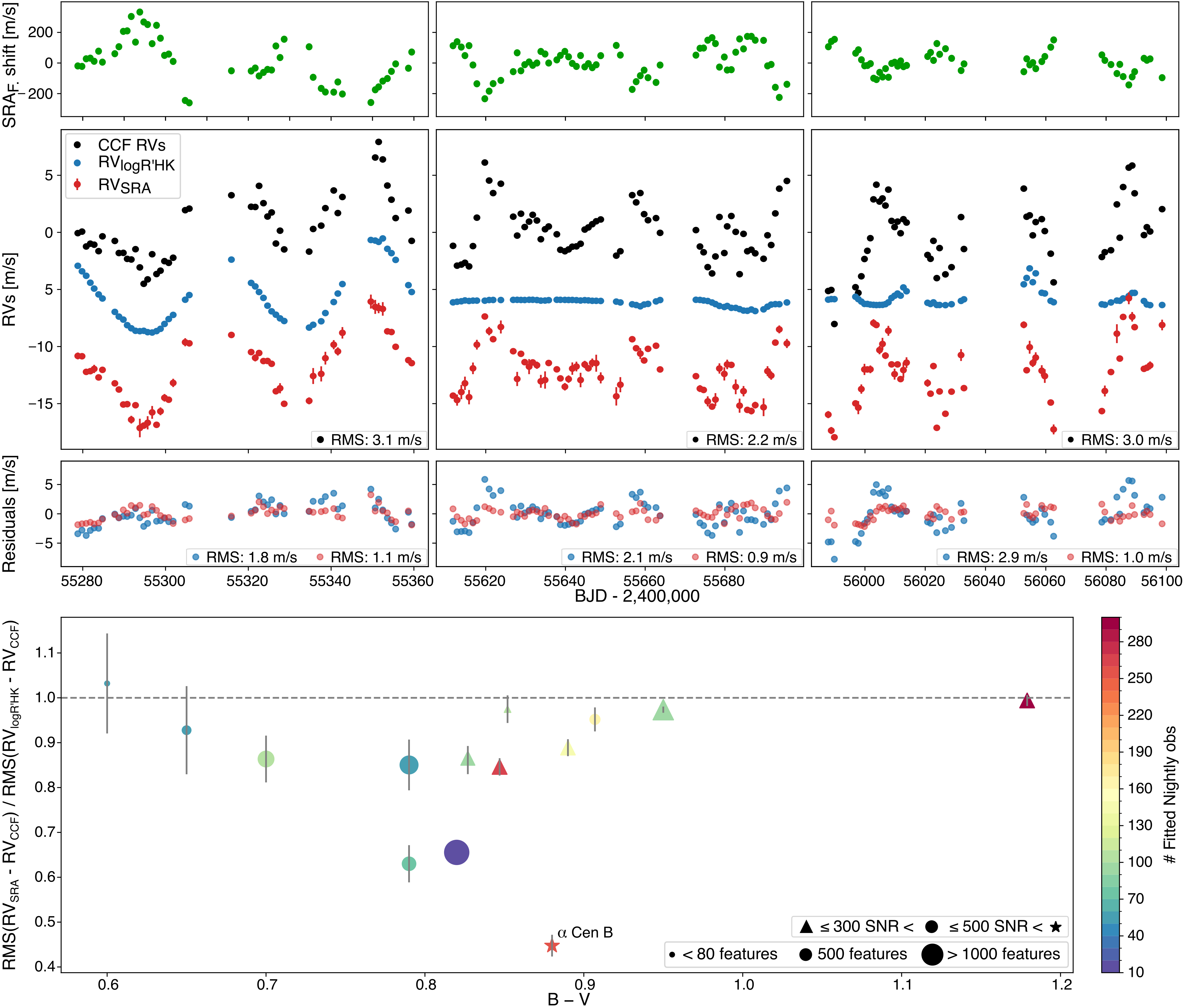}
    \caption{\label{fig:RVmodels} Comparison of the linear regression model of the CCF RVs using \logRHK\ and the SRA indicators. The upper panels show the measured SRA$_{\mathrm{Feature}}$ velocity shift values, in green, for the K1V star $\alpha$~Cen~B, where each panel focusses on a specific season. These panels serve as an example to present how the SRA$_{\mathrm{Feature}}$ shifts is measuring the effect of active regions on the surface of a star. The second row of panels show the CCF RVs, the $\mathrm{RV_{log\,R'HK}}$ and the $\mathrm{RV_{SRA}}$, in black, blue and red respectively, and offset from one another for clarity. The effects of the $\pm$~5~m~s$^{-1}$ misalignment of the spectra when using SRA are represented by the red error bars. The RMS of the raw CCF RVs are displayed in each panel. The third row of panels present, for each time series, the residuals after subtracting the $\mathrm{RV_{log\,R'HK}}$ and the $\mathrm{RV_{SRA}}$ to the CCF~RVs, in blue and red respectively. For each residual, the RMS obtained are displayed in each panel. The lower panel compares the results of the fitting for all the \nstar~stars of our sample by dividing the $\mathrm{RV_{SRA}}$ RMS by the $\mathrm{RV_{log\,R'HK}}$ RMS. Hence, a ratio above one (dashed-line) indicates that the $\mathrm{RV_{SRA}}$ based correction does not improve the fit compared to the $\mathrm{RV_{log\,R'HK}}$. Conversely, a ratio below one demonstrates that the SRA method yields a superior reduction in the CCF RV RMS. Similar to Figure~\ref{fig:corrlog}, a different shape was assigned to represent the median value of the weighted average SNR in the 55$^{\mathrm{th}}$ order for the nightly data. Additionally, the size of each point represents the number of \feat\ used in the analysis, while the colour of each star represents the number of nightly observations per star to identify possible cases of bias. Finally, the vertical grey lines represent the 0.95 confidence level of the jackknife estimate.}
\end{figure*}

We use $\alpha$~Cen~B as a case study to facilitate a more detailed discussion. This star is well suited for this illustration due to its high SNR and clear RV variations driven by rotational modulation of active regions. The upper panels of Figure~\ref{fig:RVmodels} present the measured SRA$_{\mathrm{Feature}}$ velocity shift values, in green, for three different observing seasons. These seasons were selected primarily for their dense data sampling, providing an ideal baseline to demonstrate how SRA measures the effect of active regions crossing the stellar surface. For instance, in the case of $\alpha$~Cen~B, the SRA$_{\mathrm{Feature}}$ shifts vary between $\pm$300~m~s$^{-1}$ and are clearly modulated by the rotation of the star. The second row of panels in Figure~\ref{fig:RVmodels} compare the nightly CCF RVs (black points) with the model prediction from $\mathrm{RV_{log\,R'HK}}$ (blue) and $\mathrm{RV_{SRA}}$ (red). In the first season, $\mathrm{RV_{log\,R'HK}}$ tracks the CCF RV variations well. However, its performance degrades markedly in the second and third seasons, failing to reproduce the observed RV variations. By contrast, $\mathrm{RV_{SRA}}$ clearly follows the CCF RVs for all three seasons much better than \logRHK\ does. These results can also be observed in the third row of panels, which show the residuals of each time-series (with their RMS values), after subtracting the $\mathrm{RV_{log\,R'HK}}$ and the $\mathrm{RV_{SRA}}$ from the CCF~RVs, in blue and red, respectively. The residuals of the $\mathrm{RV_{log\,R'HK}}$ are clearly noisier than those of the $\mathrm{RV_{SRA}}$, with a mean RMS of 2.3~m~s$^{-1}$ versus 1.0~m~s$^{-1}$, respectively. We note the presence of remaining structures in both residuals. These structures could be explained by the simplicity of the model applied, as well as a possible phase shift between the CCF RVs and the $\mathrm{RV_{SRA}}$, similar to the phase shifts found for the Sun between the CCF RVs and some activity indicators (see \citealt{Cameron2019}).

One potential concern with SRA-derived parameters is the effect of imperfect spectral alignment. Misalignments between the "active" and "inactive" spectra (see Section~\ref{subsec:process}) prevent identical lines (for example) from dividing out cleanly. Instead, one side of the line is over-divided, while the other side is under-divided, resulting in a distinctive trough-peak (or vice versa) `S-shaped' residual in the ratio spectra. The amplitudes and velocities of these artefacts would scale with the magnitude of the misalignment. This is of further concern if some of the measured SRA signals are driven by genuine Doppler shifts in the spectra, rather than by line-profile changes due to stellar activity. This can lead to a true misalignment of the spectra, which produces features in the SRA that, in turn, would yield a correlated signal with the CCF RVs -- but the correlation would be driven by an actual velocity displacement, not by activity-induced distortions (see Section~\ref{subsec:injection}). Under these conditions it would appear as though the SRA-derived activity indicators were tracking RV variations far better than \logRHK\ (which is insensitive to true Doppler shifts), but in such a case the apparent improvement would be an artefact of the Doppler misalignment rather than a true enhancement in the sensitivity to stellar activity.

To assess whether a true Doppler misalignment (which would generate an artificial SRA-RV correlation) was responsible for our results, we first inspected the ratio spectra. Fortunately, such misalignments would produce a set of very recognisable `S-shaped' traits as described earlier, and we are confident that the features we are seeing are not driven by incorrect alignment as such characteristic signals are absent. Moreover, such a misalignment would affect nearby spectral lines of similar depth and width equally, whereas we observe strong variability in some lines but not others that otherwise have very similar properties (see e.g. Figure~\ref{fig:StarsFlux}).

\begin{table}
    \centering
    \begin{tabular}{c|c|c|c}
    \hline
    \hline
    	&	Season 1	&	Season 2	&	Season 3		\\
    \hline							
    $\alpha_2$ (\logRHK)	    &	4.1	&	0.5	&	1.4		\\						
    \rule{0pt}{3ex}$\alpha_3$ (\logRHK$^{2}$)	    &	0.8	&	-0.3	&	1.6		\\		
    \rule{0pt}{5ex}$\beta_2$ (SRA$_{\mathrm{Feature}}$ amplitude)	    &	3.6	&	1.3	&	2.0		\\
    \rule{0pt}{3ex}$\beta_3$ (SRA$_{\mathrm{Feature}}$ shift)	    &	-2.3	&	-3.5	&	-4.7		\\
    \hline
    \hline
    \end{tabular} \\
    \caption{Fitted coefficients for the linear regression models $\mathrm{RV_{log\,R'HK}}$ and $\mathrm{RV_{SRA}}$ from Equation~\ref{eq:RVmodel} for the three seasons of $\alpha$~Cen~B presented in Figure~\ref{fig:RVmodels}. The parameters used were first normalised between $-1$ and 1 in order to compare their respective weight in the two fits. Note that the constant terms $\alpha_1$ and $\beta_1$ were excluded from this table as they only serve to absorb the systemic velocity offset.}
    \label{tab:Modelfactors}
\end{table}

Nonetheless, as a precaution, we repeated the SRA process but deliberately introduced misalignments of $\pm$~5~m~s$^{-1}$ (which is similar in magnitude to the typical CCF RV variability of the stars in our sample) when constructing the nightly ratio spectra. These misalignments were applied to the spectra before dividing the higher activity spectra by the low-activity template (see Section~\ref{subsec:process}). The results of these tests for $\alpha$~Cen~B are presented in Figure~\ref{fig:RVmodels} for the $\mathrm{RV_{SRA}}$ by the red error bars. We note that the results were largely unchanged, confirming that the observed SRA features predominantly reflect line-shape variations due to stellar activity.

Each parameter used in the linear regression model of Equation~\ref{eq:RVmodel} was normalised between $-1$ and 1 to compare their weight and determine which indicator dominates each fit. Table~\ref{tab:Modelfactors} lists these fitted coefficients from the RV modelling for $\alpha$~Cen~B. For completeness, we also provide the scale factor for the $\mathrm{RV_{log\,R'HK}}$, but no clear distinction can be observed between the two terms, except during the first season where the \logRHK\ term guides the fit by a factor four.

Concerning the $\mathrm{RV_{SRA}}$ parameters, we first note that the coefficient ($\beta_{2}$) associated with the SRA$_{\mathrm{Feature}}$ amplitude is leading the fit for the first season while the coefficient ($\beta_{3}$) associated with the SRA$_{\mathrm{Feature}}$ velocity shift dominates the second and third season. These results seem to confirm our observations from Figure~\ref{fig:corrlog} where the SRA$_{\mathrm{Feature}}$ amplitude appeared well correlated with \logRHK. Thus, from Table~\ref{tab:Modelfactors} and Figure~\ref{fig:RVmodels} we suggest that, during the first season, some specific stellar activity processes appear to be seen by both chromospheric and photospheric indicators. However, during the second and third season, a different type of activity seems to occur at the stellar surface, only detectable by the photospheric indicator SRA$_{\mathrm{Feature}}$ velocity shift.

Additionally, we also observe from Table~\ref{tab:Modelfactors} that the SRA$_{\mathrm{Feature}}$ velocity shift terms are all negative, indicating an anti-correlation with the CCF RVs. This behaviour is expected as the SRA$_{\mathrm{Feature}}$ velocity shift represents the mean-velocity of the activity-induced `bump' within the line-profile. A bump located in the blue wing of the line-profile (i.e. a blue-shifted SRA$_{\mathrm{Feature}}$ velocity) will displace the line's centre-of-light towards the red (leading to a measured red-shift in the CCF) and vice versa. This explains why $\beta_3$ is negative in all of the fits.

On the other hand, the coefficients associated with the measured SRA$_{\mathrm{Feature}}$ amplitudes are all positive, indicating that there is a predominant positive correlation with the CCF RVs. This correlation can also be easily explained as we expect that RV variations will be mainly caused by the suppression of the convective blueshift, linked to the increase of stellar activity which will result in greater deformation of the stellar lines causing stronger features in the SRA.

For completeness, Figure~\ref{fig:Allmodels} presents the results of this fitting exercise for all three seasons of $\alpha$~Cen~B, using both first- and second-order polynomial models for comparison. Each panel displays the CCF RVs alongside fits based on \logRHK, SRA${_\mathrm{Feature}}$ amplitudes, and SRA${_\mathrm{Feature}}$ velocity shifts, with the corresponding Pearson’s $R$ correlation values given. Additionally, Table~\ref{tab:2} compares the results of several combinations of different activity indicators. We tested \logRHK, CCF FWHM, and CCF BIS with Equation~\ref{eq:RVmodel} and compare the resulting RMS values for the three seasons of $\alpha$~Cen~B. While the choice of the indicators can lead to different RMS value, in all cases, none of the standard activity indicator combinations reached the RMS levels achieved by the $\mathrm{RV_{SRA}}$. These results reinforce our conclusions from Figure~\ref{fig:RVmodels}. The SRA$_{\mathrm{Feature}}$ amplitudes and shifts better represent the variations in the CCF RVs due to stellar activity than \logRHK. In particular, the SRA$_{\mathrm{Feature}}$ velocity shifts consistently track the CCF RV variations across all three seasons with Pearson's~$R>0.8$. This is consistent with our scalar/vector framing of these different activity indicators: \logRHK\ and the SRA$_{\mathrm{Feature}}$ amplitudes are `scalar-like' activity measures that can align with RV variability when their phase matches, but the SRA$_{\mathrm{Feature}}$ velocity shifts provide a complementary `directional' component that appears more able to maintain predictive power even when the scalar-like terms lose coherence.

\subsection{From case study to sample-wide linear RV modelling}\label{subsec:allstars}
In this Section we report on the more general results of the same linear regression model approach applied to all \nstar~stars in our sample. For each star we followed the same process as before, using the two models outlined in Equation~\ref{eq:RVmodel} to fit to the CCF RVs. We then compared the RMS of the residuals of these fits based on \logRHK\ and the SRA-derived summary metrics. The lower panel of Figure~\ref{fig:RVmodels} shows the ratio of the RMS from the SRA-based fit to that from the \logRHK-based fit. As in Figure~\ref{fig:corrlog}, the marker size reflects the number of \feat\ used in the analysis, the marker colour represents the number of nightly observations used in the modelling per star, and the marker shape indicates the median for the nightly weighted average of the SNR in the 55$^{\mathrm{th}}$ order. Additionally, the point showing $\alpha$~Cen~B is highlighted in Figure~\ref{fig:RVmodels}. Similar to Figure~\ref{fig:corrlog}, the vertical grey lines were added to represent the 0.95 confidence level of the jackknife estimate. We note that one star does not have any variability measurement as all the fitted nights in the model came from the same season.

\begin{figure*}
    \centering
    \includegraphics[width=1\textwidth]{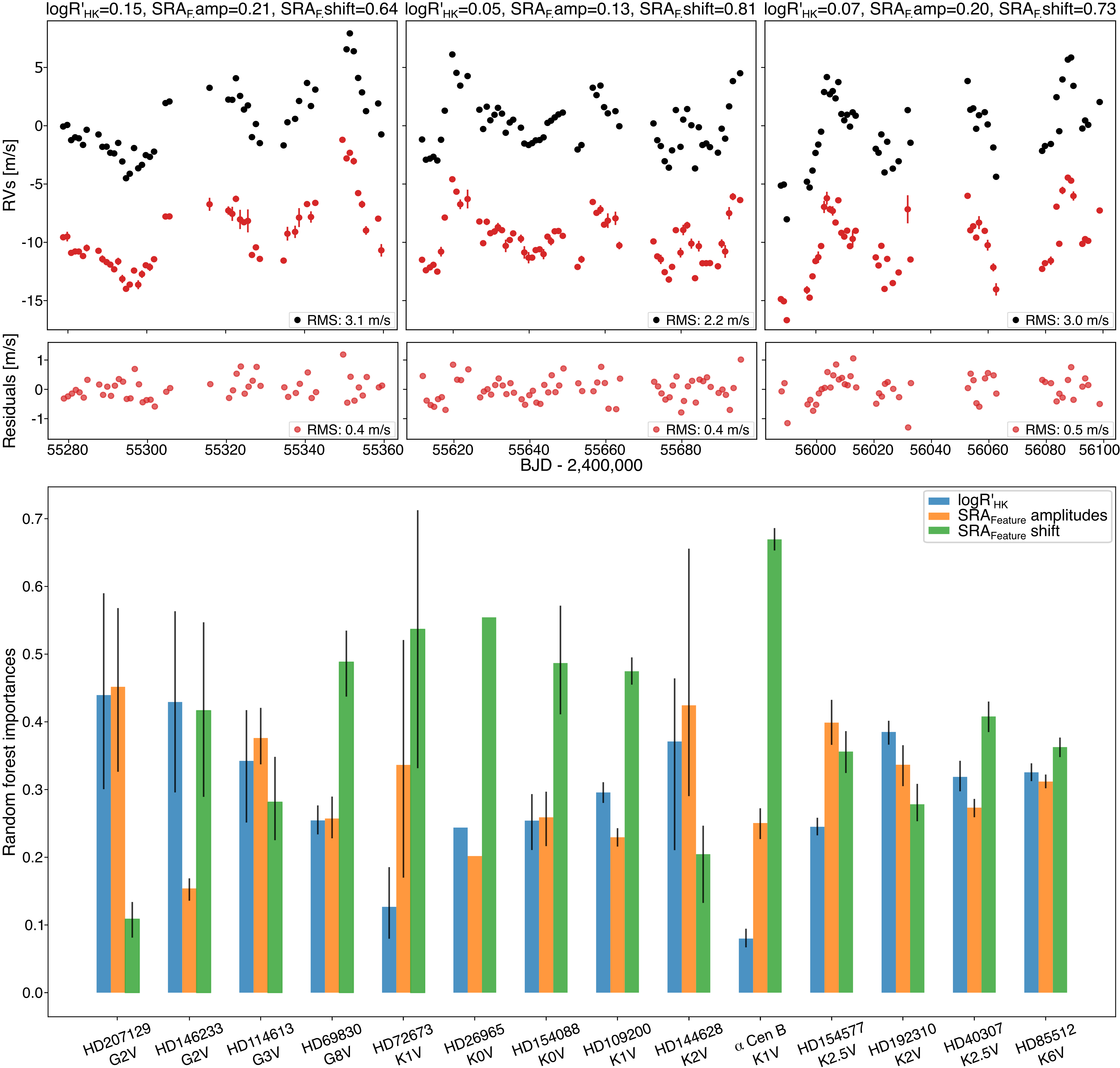}
    \caption{\label{fig:RVmodel_RF} Probing the importance of \logRHK, SRA$_{\mathrm{Feature}}$ amplitudes and SRA$_{\mathrm{Feature}}$ shifts using random forest regression. The upper panels show the CCF RVs and the results of the random forest regression for $\alpha$~Cen~B in black and red, respectively, and offset from one another for clarity. Similar to Figure~\ref{fig:RVmodels} each panel represents a different season, the effects of the $\pm$~5~m~s$^{-1}$ misalignment of the spectra when using SRA are represented by the red error bars and the RMS of the raw CCF RVs are displayed in each panel. The importance of the three indicators used are shown at the top of each panel. The middle panels present for each time series the residuals after subtracting the random forest regression prediction to the CCF RVs. The RMS obtained from the residuals are displayed in each panel. The lower panel presents the importance of \logRHK, SRA$_{\mathrm{Feature}}$ amplitudes and SRA$_{\mathrm{Feature}}$ shifts for the \nstar~stars, in blue, orange and green, respectively. The ordering of the stars follows the same order as Figure~\ref{fig:StarsFlux} and Table~\ref{tab:1} (i.e. ordered by $B - V$). Since this analysis was done on a season by season basis, we applied a weighted average, where the weight was the number of nightly spectra per season. Finally, the vertical grey lines represent the 0.95 confidence level of the jackknife estimate.}
\end{figure*}

By dividing the $\mathrm{RV_{SRA}}$ RMS by the $\mathrm{RV_{log\,R'HK}}$ RMS, we observe in Figure~\ref{fig:RVmodels} that most of the stars lie below the 1.0~dashed-line, indicating that $\mathrm{RV_{SRA}}$ yields smaller residuals (and is better tracking the CCF RV variations) than $\mathrm{RV_{log\,R'HK}}$. We find some dependence of the results on the number of features and the weighted average SNR appears to matter significantly, with (effective) SNRs above 300 a practical threshold needed for the SRA-derived activity indicators to outperform \logRHK\ in this modelling. A SNR of $\sim$300 represents what can be achieved in a single relatively well-exposed HARPS spectrum, indicating that the useful application of SRA is well within practical reach of current spectrographs.

Additionally, we measured the SRA$_{\mathrm{Feature}}$ velocity shift variations for all \nstar~stars in our sample. The results are displayed as a box plot in Figure~\ref{fig:SRAvar}, where the bottom of the box, the black horizontal line, and the top of each box correspond to the 25$^{\mathrm{th}}$, 50$^{\mathrm{th}}$, and 75$^{\mathrm{th}}$ percentiles, respectively, colour coded by the $B - V$ colour index. We note that the dispersion is larger for the hottest stars, indicating that the SRA$_{\mathrm{Feature}}$ velocity shift is primarily driven by the stellar rotation, as hotter stars are typically faster rotators. However, other parameters, such as the inclination of the star may also influence the spread of the SRA$_{\mathrm{Feature}}$ shifts.

Similarly to Section~\ref{subsec:alphacenb}, we tested several combinations of activity indicators (i.e. \logRHK, CCF FWHM and CCF BIS) for all \nstar~stars in our sample. The results are summarized in Table~\ref{tab:3}. As seen for the case of $\alpha$~Cen~B, while the choice of activity indicator can lead to different RMS values, the results obtained using the $\mathrm{RV_{SRA}}$ (i.e. last column of Table~\ref{tab:3}) yield the lowest RMS values for the majority of the stars and configurations studied. In only two cases (highlighted in red), were we able to reach a RMS value slightly below the one obtained using $\mathrm{RV_{SRA}}$.

In conclusion, we have shown that for stars observed with high SNR, the SRA indicators can better mitigate the effects of stellar activity in RV measurements compared to classical and widely used activity proxies. This confirms our scalar/vector analysis above with $\alpha$~Cen~B, where we showed that SRA$_{\mathrm{Feature}}$ velocity shifts, which act as a `directional' component, can provide better predictive power in comparison to the `scalar' ones. However, it is important to note that the SRA-derived indicators can be degraded by significant Doppler shifts and/or misaligned spectra (e.g. planets, barycentric errors, binaries, see below Section~\ref{subsec:injection}). Hence, care should be taken when applying SRA as the SRA indicators could pick up both true activity-driven line distortion and artefacts from misalignments.

\subsection{Further benchmarking of SRA-derived activity indicators with random forest regression} \label{subsec:randomforest}
In addition to the linear regression model analysis, we compared the performance of \logRHK, CCF FWHM and CCF BIS with the new SRA-derived indicators (i.e. SRA$_{\mathrm{Feature}}$ amplitudes and SRA$_{\mathrm{Feature}}$ velocity shifts) for mitigating astrophysical noise using random forest regression. Random forests are an ensemble learning method that combine multiple decision trees on samples of the data (see \citealt{Breiman2001} for a detailed description). In this study, we only evaluate which of the three indicators of each set is the most influential in the random forest regression when predicting the stellar RV variations. We therefore used the random forest importance function to quantify the contribution of each indicator.

Figure~\ref{fig:RVmodel_RF} presents the results of the random forest regression and, as above, we use the case of $\alpha$~Cen~B as an illustration before reporting on the sample-wide results. As previously discussed, we adopted \logRHK\ as our benchmark, and thus only the results obtained among the \logRHK, the SRA$_{\mathrm{Feature}}$ amplitudes and SRA$_{\mathrm{Feature}}$ velocity shifts are shown in Figure~\ref{fig:RVmodel_RF}. We discuss below the results obtained using the other activity indicators. The upper panels of Figure~\ref{fig:RVmodel_RF} compare the nightly CCF RVs (black) with the random forest predictions (red) for the three seasons of $\alpha$~Cen~B. The importance of the three indicators are shown at the top of each panel. Since the goal was to fit the CCF RVs using the activity indicators, 90\% of the data were used in the training of the random forest. For $\alpha$~Cen~B we find that \logRHK\ have importance values that only contribute weakly ($\leq\ 0.15$), whereas the SRA$_{\mathrm{Feature}}$ velocity shifts dominate the importance across all three seasons ($\geq\ 0.64$). These results reinforce the conclusion that velocity information contained in the SRA-derived features extracted from photospheric lines appears to be a powerful predictor of activity-induced RV variability. As above, we assessed the effects of spectral misalignment by introducing $\pm$~5~m~s$^{-1}$ shifts when creating the nightly ratio spectra. The resulting fits, represented by the red error bars, are also similar to what we found above (see Figure~\ref{fig:RVmodels}), and reaffirms that activity-induced line-shape variability still dominates the results, with only minor sensitivity to the imposed velocity offsets. The middle panels show the residuals between the CCF RVs and the results of the random forest regression.

Turning now to the full stellar sample, the lower panel of Figure~\ref{fig:RVmodel_RF} summarises the random forest importance of each of the three indicators (\logRHK, SRA$_{\mathrm{Feature}}$ amplitude, and SRA$_{\mathrm{Feature}}$ velocity shift, in blue, orange and green, respectively). Similar to the approach used in Section~\ref{subsec:allstars}, the random forest regression was applied to each season independently to ignore some of the effects of long-term variations. The determined importances of each star were then weighted using the number of fitted nights per season. Similar to Figures~\ref{fig:corrlog}~\&~\ref{fig:RVmodels}, the vertical grey lines represent the 0.95 confidence level of the jackknife estimate.

From Figure~\ref{fig:RVmodel_RF}, we observe that the SRA$_{\mathrm{Feature}}$ velocity shift has the highest importance for the majority of the stars in our sample (eight out of \nstar), whereas the contribution of \logRHK\ barely dominates for only two stars. The same results can also be seen in Table~\ref{tab:4}, which displays the importances obtained when using other activity indicators. We can observe that the contribution of the CCF BIS is never greater than that of the SRA indicators. For the CCF FWHM, four stars exhibit higher importances and these cases are highlighted in red in Table~\ref{tab:4}. Upon investigation, we found that these four stars correspond primarily to targets with lower SNR and/or poor sampling. This can prevent features from being properly detected, resulting in a low number of \feat\ detected and higher noise levels in our measured SRA$_{\mathrm{Feature}}$ velocity shift. This highlights the need for high-SNR (above 300) and high-cadence observations, as discussed in Section~\ref{subsec:allstars} and in Figure~\ref{fig:RVmodels}. These observations confirm our previous results from Section~\ref{subsec:allstars}. SRA-derived indicators (and, in particular, the SRA$_{\mathrm{Feature}}$ velocity shift information) show promise as a novel and powerful photospheric tracer of stellar activity-induced RV variability, offering a more reliable proxy than classical activity indicators across the majority of the stellar sample studied here.

\subsection{Planetary injection recovery tests} \label{subsec:injection}
A final test to benchmark the use of SRA and its \feat\ was to measure their capacity to retrieve fake injected planetary signals. The first goal of this test was to study the limits of SRA and to measure the impact that missed or poorly removed planetary signals can have on the \feat. Additionally, the second goal of this test was to compare which of the two linear regression models between the $\mathrm{RV_{log\,R'HK}}$ and the $\mathrm{RV_{SRA}}$ better retrieves the injected signals. Using $\alpha$~Cen~B, we performed five tests where we injected a planetary signal at an orbital period of 100~d with a semi-amplitude of 10, 5, 2, 1 and 0.5~m~s$^{-1}$. 

\begin{figure*}
    \centering
    \includegraphics[width=1\textwidth]{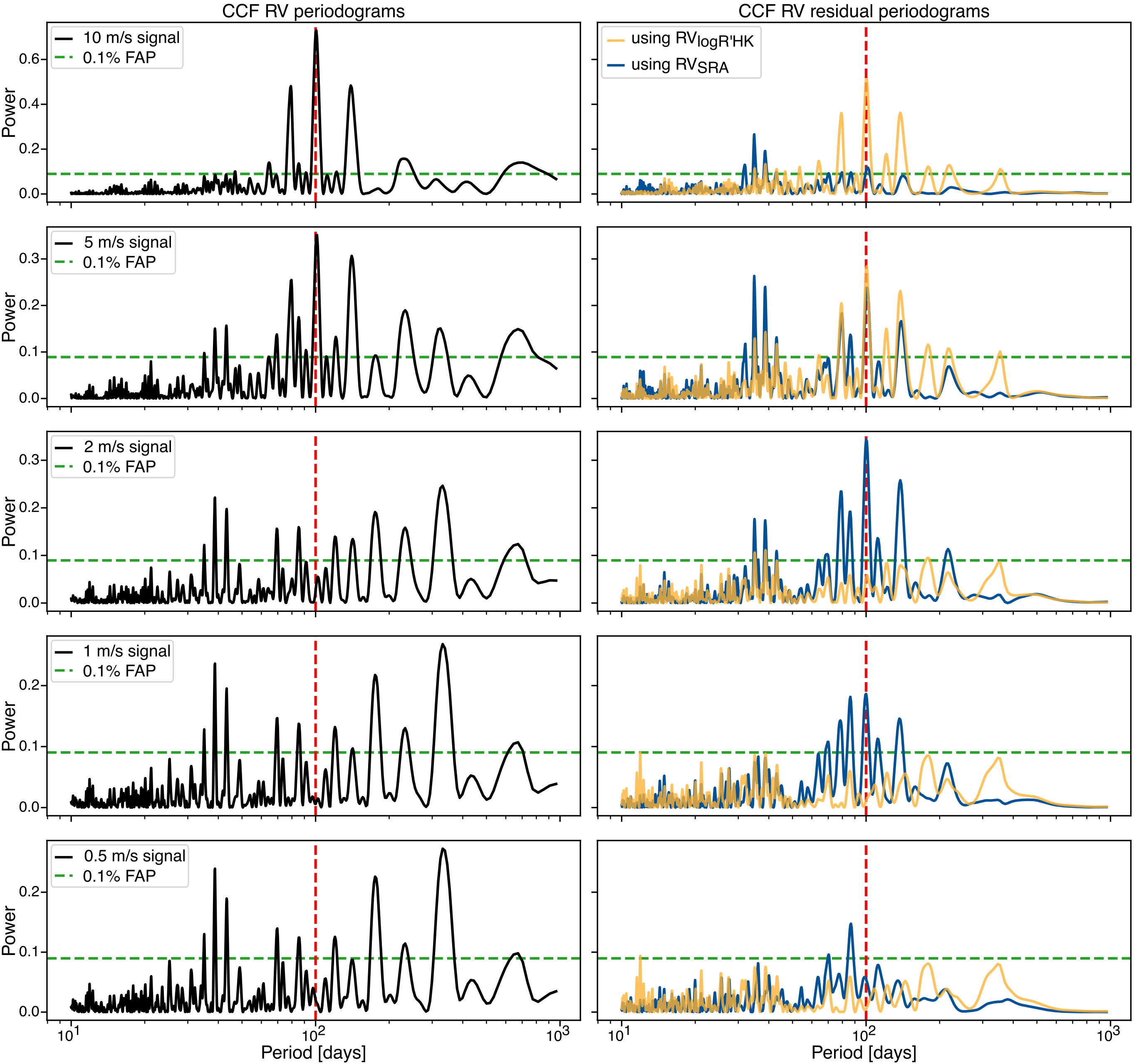}
    \caption{\label{fig:Injection} Comparison of different planetary signals injection and their recovery in the residuals. The left panels represent, in black, the CCF RV periodograms of $\alpha$~Cen~B after injecting five planetary signals (one signal per panel), with a semi-amplitude of 10, 5, 2, 1 and 0.5~m~s$^{-1}$ at an orbital period of 100~d. The fake planetary signal is marked with a vertical red dashed line and the 0.1\% FAP level is shown with a horizontal green dashed line. The right panels represent the CCF RV residual periodograms after removing the $\mathrm{RV_{log\,R'HK}}$ and the $\mathrm{RV_{SRA}}$ from the CCF RVs for the five cases, in yellow and blue, respectively. Except in the cases where the injected signal are important enough to be directly seen by eye in the periodograms of the CCF RVs (at 10 and 5~m~s$^{-1}$), we observe that the residuals obtained using \logRHK\ do not help in retrieving the planetary signal. However, the residuals obtained using the SRA$_{\mathrm{Feature}}$ amplitudes and shifts allow to retrieve the hidden injected signal at 2 and 1~m~s$^{-1}$, showing the benefit of the SRA activity indicators.}
\end{figure*}

The left panels of Figure~\ref{fig:Injection} present, in black, the five CCF RV periodograms of $\alpha$~Cen~B after the injection of the fake signal, marked with a vertical red dashed line. In each panel, the 0.1\% FAP level is shown with a horizontal green dashed line. The right panels of Figure~\ref{fig:Injection} present the CCF RV residual periodograms after removing the $\mathrm{RV_{log\,R'HK}}$ and the $\mathrm{RV_{SRA}}$ from the CCF RVs for the five cases, in yellow and blue, respectively. For the 10 and 5~m~s$^{-1}$ cases, the injected signals are strong enough to be seen directly by eye in the CCF RV periodograms. Thus, the residual periodograms do not provide much additional information. We still note that at 10~m~s$^{-1}$, the planetary signal has disappeared from the $\mathrm{RV_{SRA}}$ residuals. This can be explained by the fact that the fake planetary signal is so strong that it takes over the \feat. Thus, SRA mainly tracks the shift variations of the lines caused by the planetary signal, represented in the residuals by `S-shaped' features, rather than the features from stellar activity. For the 5~m~s$^{-1}$ case, while we are able to detect the injected signal in the $\mathrm{RV_{SRA}}$ residuals, the recovered signal also seems slightly absorbed by the SRA technique, in comparison with the $\mathrm{RV_{log\,R'HK}}$ residuals. These tests serve as a reminder that SRA should not be used blindly without removing the important RV shift signals present in the data.

In the 2 and 1~m~s$^{-1}$ cases, the injected signals do not produce discernible peaks in the CCF RV periodograms. Thus, in order to retrieve the injected signal, we need to reduce the noise coming from stellar activity. For both cases, the $\mathrm{RV_{log\,R'HK}}$ failed to retrieve the planet in the residuals, as no peak can be seen in the GLS periodogram at the expected period for the orange curves. However, the signal of the fake planet can be retrieved in the residuals using the $\mathrm{RV_{SRA}}$, as can be seen in Figure~\ref{fig:Injection}, where a strong peak is observed at the expected period in the GLS periodogram for the green curves. This shows that the SRA activity indicators were able to remove the noise coming from stellar activity in order to detect the hidden planet. Finally, at 0.5~m~s$^{-1}$, the signal cannot be retrieved using the $\mathrm{RV_{log\,R'HK}}$ or the $\mathrm{RV_{SRA}}$, which is not surprising considering the faint signal and the simplicity of our linear regression models.

From these simple planetary injection recovery tests, we were able to show the ability of SRA to remove stellar noise over more conventional activity indicators such as \logRHK. These tests confirm once again that the velocity information contained in the \feat\ can be a powerful predictor of activity-induced RV variability. These tests were also useful to determine the limits of SRA and to observe how the SRA-derived indicators can be degraded by significant Doppler shifts.

\section{Summary}\label{sec:conclusion}

In this paper, we have presented a proof-of-concept demonstration showing that SRA indicators can be used to better trace RV variability. This technique allowed us to analyse the effects that stellar activity has on the stellar spectra for \nstar~relatively quiet and well observed G- and K-type stars. By dividing high-activity spectra by a low-activity template, we showed that SRA can be used to isolate activity-driven changes directly in the stellar photospheric absorption lines where RVs are measured.

We first described the SRA process and the methods used to identify these \feat. From our data, hundreds of \feat\ were found for the different stars used in our sample. For this proof-of-concept, we have distilled the information in the SRA ratio spectra into two global metrics measuring the amplitude and shift variations of the features over nightly and seasonal data. By analysing the variations of these \feat, some correlations were found with the variation of the stellar activity using the chromospheric indicator \logRHK\ for the \nstar~stars:
\begin{itemize}
  \item The SRA$_{\mathrm{Feature}}$ amplitudes are well correlated with \logRHK. This is true over both the short and long time scales. In addition, the correlation was found to decrease for hotter stars. We theorise that this trend is probably due to the greater difference in atmospheric height between the photosphere and the chromosphere for the earlier-type stars. These results show that the activity-sensitive feature amplitudes can act as a tracer of the overall activity level of a star.
  \item The SRA$_{\mathrm{Feature}}$ velocity shifts show no correlation with \logRHK\ over short (stellar rotation periods) time scales. This is expected as the SRA$_{\mathrm{Feature}}$ shifts track the velocity of active regions on the stellar surface, which differs from the activity level of a star. However, surprisingly, a positive correlation is observed between the SRA$_{\mathrm{Feature}}$ shifts and \logRHK\ over long (activity cycle length) time scales for cooler stars. We speculate that this correlation may be due to the suppression of the convective blueshift increasing as the star becomes more active as it progresses through its magnetic cycle.
\end{itemize}

Following our analysis of these SRA metrics, we benchmarked the SRA indicators against \logRHK, CCF FWHM and CCF BIS to evaluate their utility in mitigating astrophysical noise in RV measurements. Using both the SRA$_{\mathrm{Feature}}$ amplitudes and shifts, we fitted the CCF RVs on shorter time scales and compared these results against those obtained using classical stellar activity proxies. By applying these fits to the different stars in our sample, we find that SRA-derived metrics capture RV variability better than other activity indicators by up to a factor of two. In particular, the SRA$_{\mathrm{Feature}}$ shifts appear to yield strong predictive power. We hypothesise that the improvements shown are primarily driven by two aspects: 
\begin{itemize}
  \item We derive activity indicators from specific, targeted photospheric lines selected for their high sensitivity to stellar activity. This tailored selection provides greater capability to stellar activity than chromospheric indicators (e.g. \logRHK) or global CCF-derived proxies (e.g. FWHM and BIS) which can dilute the localised signatures of stellar activity.
  \item The velocity information contained within the SRA metrics offers a more direct link to the sources of RV perturbations. SRA acts like a vector quantity, encoding both the magnitude (via SRA$_{\mathrm{Feature}}$ amplitudes) and the direction (via SRA$_{\mathrm{Feature}}$ shifts), which provide a new potentially diagnostic tracking the location of active regions on the stellar surface.
\end{itemize}

We note that a high effective SNR, above 300 per night, seems necessary to accurately measure the SRA parameters. Additionally, we showed that while we are relatively insensitive to small errors in the alignment process that SRA requires, Doppler shifts and/or misaligned spectra can still degrade the SRA-derived indicators. We also employed random forest regression to measure the relative importances of \logRHK, CCF FWHM, CCF BIS, the SRA$_{\mathrm{Feature}}$ amplitudes, and the SRA$_{\mathrm{Feature}}$ shifts when predicting the CCF RVs. Similarly to the fits, the random forest regression showed that in most cases, it was the SRA$_{\mathrm{Feature}}$ shifts that were the most important indicator in terms of tracking the RV variations induced by stellar activity. Finally, planetary injection recovery tests were performed. These tests confirmed the ability of SRA to further improve stellar noise mitigation over more conventional activity indicators such as \logRHK, and to study the limits of SRA in the presence of significant and uncorrected Doppler shifts.

We also note that across our various tests, the SRA indicators perform particularly well at tracking stellar activity variations in the case of $\alpha$~Cen~B compared to the other stars. This could be explained by the very high SNR of the nightly averaged spectra (i.e. SNR above 600) for this target. Additionally, $\alpha$~Cen~B has one of the best data sampling among the \nstar~stars in our sample, allowing us to accurately track the signals driven by stellar activity during clear rotational modulation.

It should be stressed that we have limited our analysis in this paper using only the amplitudes and shifts of the features. However, there is a wealth of information contained within the SRA results, including different line morphologies and more complex information that is lost by reducing the measurements to the two global metrics. To highlight this aspect, the tests presented in this paper were repeated by combining the nightly and seasonal SRA$_{\mathrm{Feature}}$ amplitudes and velocity shifts of each feature with a simple median instead of the weighted average using the fitting error (see Section~\ref{sec:SRAindicators}) and better results were obtained. For instance, all the stars presented in this paper were found to have a lower RMS ratio than those presented in Figure~\ref{fig:RVmodels}. Similarly, instead of eight stars being dominating by the importances of the random forest with the SRA$_{\mathrm{Feature}}$ shifts, that number increased to eleven. Thus, using a crude way such as a simple median when combining the SRA information of each feature seems to enhanced our predictive capability of the variations of the CCF RVs, over more `sophisticated' method such as the weighted average. The reasons behind the difference in these results are not fully understood at the moment. These differences could come from an issue in the Gaussian fitting applied to each feature, resulting in a degradation of the SRA parameters obtained when using the weighted average. Another hypothesis could be explained by the various and rich information present in the \feat. For instance, while stronger features may vary more in amplitude, their velocity shift could be diminished due to some physical difference in the impact of stellar activity for some stellar lines. Yet, these stronger features have higher weights in the weighted average, which would impact the obtained SRA$_{\mathrm{Feature}}$ shifts, and therefore, degrade the results observed here. This shows that the diverse range of morphologies and temporal evolution of these \feat\ need further investigations.

To that purpose, we have provided an atlas with the high activity averaged SRA spectra for the \nstar~stars used in this study. This atlas also contains, for each star, the nightly and seasonal SRA spectra per order, with their respective errors, wavelength, BJD and \logRHK. Additionally, all of the SRA-derived features are also provided in that atlas, with their wavelength, amplitude, width, and potentially associated line obtained from the VALD line list. We hope that this atlas will prove useful for complementary studies, such as \feat\ identification, line-by-line analysis, or magnetohydrodynamic simulations of active stellar photospheres.
For example, these results can be further validated by comparing to high-resolution spectra from MURaM (see e.g. \citealt{Beeck2015b, Sowmya2026}) and performing similar SRA analysis on simulated lines at different activity levels, to confirm that SRA$_{\mathrm{Feature}}$ shifts can indeed track the location and strength of active regions, and to what extent. SRA is a promising technique for improving astrophysical noise mitigation, and future papers will provide a more in-depth analysis on these additional metrics to improve our knowledge on stellar activity and its impact on the RVs measurement.

\section*{Acknowledgments}
\label{sec:acknow}
J.C.C., C.A.W. and E.dM. would like to acknowledge support from the UK Science and Technology Facilities Council (STFC, grant number ST/X00094X/1). K.L.H. is supported by a UK Science and Technology Facilities Council (STFC) Studentship (ST/X508706/1).
Y.C.U. would like to acknowledge support from the UK Science and Technology Facilities Council (STFC, grant number ST/W000989/1). D.S.Y. would like to acknowledge support from a UK Science and Technology Facilities Council (STFC) Studentship (ST/Y509231/1).
The Flatiron Institute is a division of the Simons Foundation. 
This research has made use of the SIMBAD database, operated at CDS, Strasbourg, France.
This work has made use of the VALD database, operated at Uppsala University, the Institute of Astronomy RAS in Moscow, and the University of Vienna. Based on publicly available data products from observations by ESO telescopes at the La Silla Paranal Observatory under programme IDs: 72.C-0488, 72.C-0513, 74.C-0012, 60.A-9036, 76.C-0878, 77.C-0530, 77.C-0364, 78.C-0833, 79.C-0681, 192.C-0852, 102.C-0584, 103.C-0206, 106.215E.002, 105.20AK.002, 183.D-0729, 183.C-0972, 188.C-0265, 185.D-0056, 196.C-1006, 198.C-0836, 99.C-0491, 100.D-0444, 101.C-0275, 103.D-0445, 1102.C-0923, 73.C-0784, 100.C-0487, 91.C-0936, 81.C-0842, 72.C-0096, 73.D-0038, 74.D-0131, 75.D-0194, 76.D-0130, 78.D-0071, 79.D-0075, 80.D-0086, 81.D-0065, 96.C-0499, 60.A-9709, 83.C-1001, 84.C-0229, 85.C-0318, 86.C-0230, 87.C-0990, 88.C-0011, 89.C-0050, 90.C-0849, 92.C-0579, 93.C-0062, 94.C-0797, 95.C-0040, 96.C-0053, 97.C-0021, 82.C-0315, 102.C-0525, 110.24BB.001, 60.A-9109

\section*{Data Availability}
The datasets were derived from sources in the public domain: [ESO~archive: \url{http://archive.eso.org/wdb/wdb/adp/phase3_main/form}, and SIMBAD: \url{http://simbad.u-strasbg.fr/simbad/}]. The relevant ESO programme IDs are listed in the Acknowledgments. 
The spectral ratio analysis (SRA) dataset underlying this article is publicly available on Zenodo at \url{https://doi.org/10.5281/zenodo.18231181}. The specific FITS structures, nightly/seasonal data orders, and text-file feature atlases are preserved under citation reference \citep{Costes2026_dataset}.


\bibliographystyle{mnras}
\bibliography{paper} 



\appendix
\section{Properties of the stars used}

\begin{table*}
    \centering
    \setlength{\tabcolsep}{5pt}
    \begin{tabular}{l|c|c|c|c|c|c|c|c|c|c|c|c|c|c}
    \hline
    \hline
    Name	&	Spectral	&	\# of	&	\# of	&	\# of	&	Median	&	Time	&	$v \sin i$		&	Median	&	$\Delta_{\mathrm{act}}$	&	$B - V$	&	Seasonal	&	Nightly	&	\# of	\\
	&	type	&	spectra	&	active	&	active	&	 \# of nights	&	range	&	[km s$^{-1}$]		&	\logRHK	&	 	&		&	averaged	&	averaged	&	features	\\
	&		&		&	nights	&	seasons	&	per season	&	[days]	&			&		&	 	&		&	SNR	&	SNR	&		\\
	\hline
    HD207129	&	G2V	&	338	&	67	&	5	&	16	&	3997	&	3.8	$^1$	&	-4.91	&	0.07	&	0.60	&	1399	&	342	&	81	\\
HD146233	&	G2V	&	4987	&	82	&	10	&	7	&	3924	&	3.7	$^1$	&	-4.93	&	0.09	&	0.65	&	996	&	371	&	261	\\
HD114613	&	G3V	&	638	&	109	&	7	&	12	&	4112	&	4.3	$^1$	&	-5.11	&	0.12	&	0.70	&	1403	&	432	&	631	\\
HD69830	&	G8V	&	623	&	100	&	11	&	15	&	4205	&	2.7	$^1$	&	-4.97	&	0.08	&	0.79	&	1345	&	359	&	755	\\
HD72673	&	K1V	&	438	&	88	&	6	&	9	&	3456	&	2.0	$^1$	&	-4.92	&	0.07	&	0.79	&	1009	&	321	&	506	\\
HD26965	&	K0V	&	483	&	53	&	12	&	5	&	4100	&	2.0	$^1$	&	-4.97	&	0.16	&	0.82	&	989	&	458	&	1149	\\
HD154088	&	K0V	&	264	&	104	&	7	&	22	&	3317	&	3.2	$^1$	&	-5.06	&	0.05	&	0.83	&	1363	&	279	&	532	\\
HD109200	&	K1V	&	716	&	270	&	7	&	32	&	3656	&	1.7	$^2$	&	-4.95	&	0.10	&	0.85	&	1588	&	253	&	578	\\
HD144628	&	K2V	&	247	&	118	&	6	&	8	&	3657	&	2.5	$^2$	&	-4.94	&	0.07	&	0.85	&	728	&	246	&	155	\\
$\alpha$~Cen~B	&	K1V	&	16678	&	279	&	8	&	42	&	3028	&	1.1	$^3$	&	-4.92	&	0.14	&	0.88	&	3723	&	623	&	600	\\
HD154577	&	K2.5V	&	538	&	158	&	7	&	22	&	3942	&	3.1	$^2$	&	-4.89	&	0.08	&	0.89	&	1105	&	235	&	565	\\
HD192310	&	K2V	&	1514	&	178	&	8	&	48	&	3991	&	2.0	$^1$	&	-4.98	&	0.16	&	0.91	&	2365	&	378	&	327	\\
HD40307	&	K2.5V	&	425	&	157	&	9	&	15	&	4138	&	2.2	$^1$	&	-4.91	&	0.15	&	0.95	&	965	&	257	&	907	\\
HD85512	&	K6V	&	1021	&	354	&	7	&	50	&	3389	&	2.2	$^2$	&	-4.97	&	0.19	&	1.18	&	1668	&	241	&	574	\\
    \hline
    \hline
    \end{tabular} \\
    $v \sin i$ references: $^{1}$\cite{Perdelwitz2024}, $^{2}$\cite{Soto2018}, $^{3}$\cite{Herrero2012}.
    \caption{Table of the \nstar~stars used in this study, ordered by $B - V$. The spectral type and $B - V$ values were taken from SIMBAD. $\Delta_{\mathrm{act}}$ represents the difference between the highest and lowest activity level of a star measured using the seasonally binned \logRHK\ values. We also note that the number of nightly and seasonal spectra reported in this table corresponds only to the high-activity ones (i.e. spectra with a \logRHK\ higher than the one from the low-activity template) used in this analysis.}
    \label{tab:1}
\end{table*}

\section{CCF RVs linear regression model}
\begin{figure*}
    \centering
    \includegraphics[width=1\textwidth]{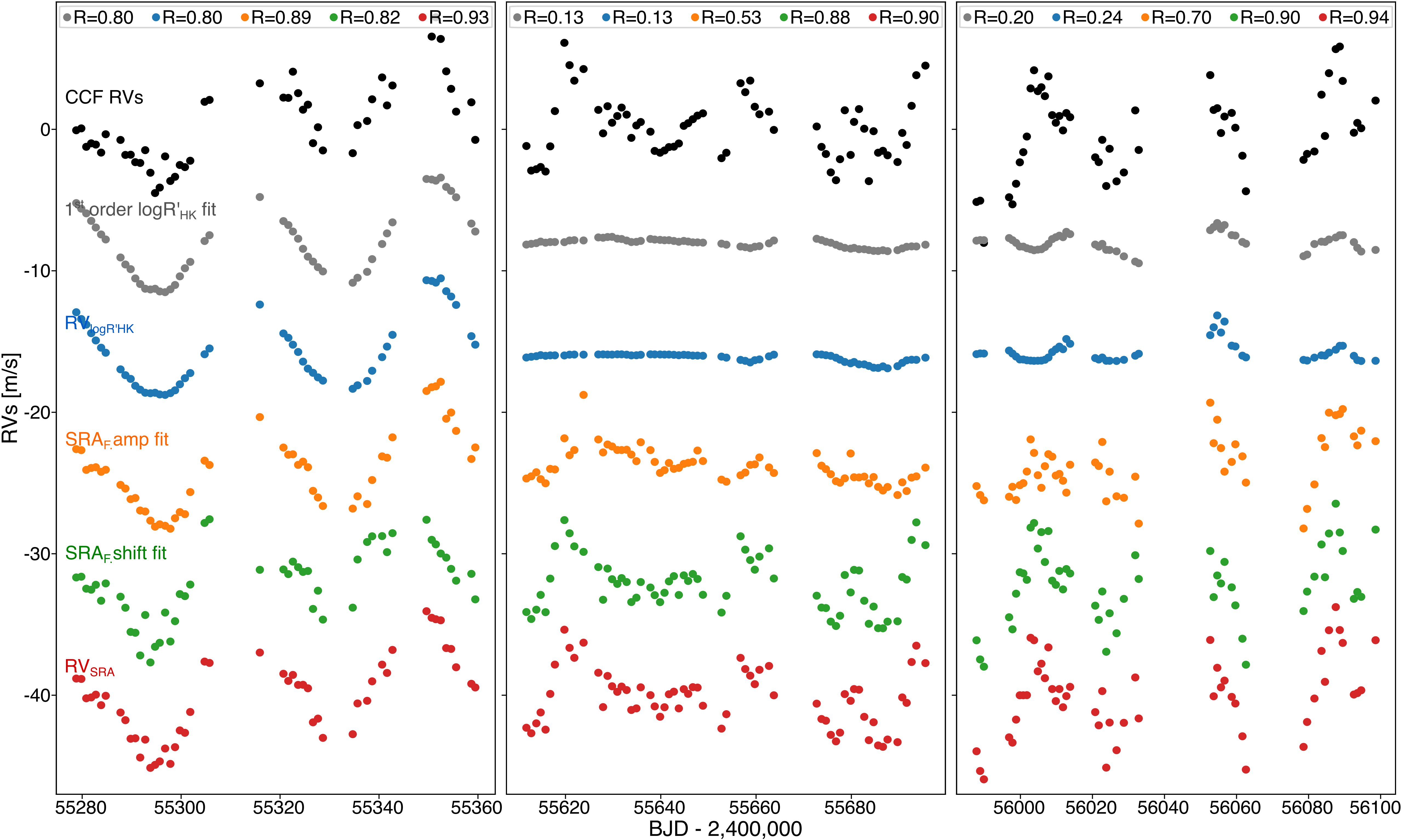}
    \caption{\label{fig:Allmodels} An overall comparison between the CCF RVs, in black, and the fits using activity indicators for $\alpha$~Cen~B. Similar to Figure~\ref{fig:RVmodels}, each panel focuses on a specific season. The fits using a first order polynomial for \logRHK, SRA$_{\mathrm{Feature}}$ amplitudes and SRA$_{\mathrm{Feature}}$ shifts are shown, respectively, in grey, orange and green. The fits for $\mathrm{RV_{log\,R'HK}}$ and $\mathrm{RV_{SRA}}$, as described in Equation~\ref{eq:RVmodel}, are shown in blue and red, respectively. At the top of each panel, the Pearson's~$R$ correlations between the CCF RVs and the different activity fits are shown.}
\end{figure*}


\section{Comparing the measured RMS using different stellar activity indicators for $\alpha$~Cen~B}
\begin{table*}
    \centering
    \begin{tabular}{l|l|c|c|c}
    \hline
    \hline
    First	&	Second	&	RMS - Season 1	&	RMS - Season 2	&	RMS - Season 3\\
	parameter &	parameter &	[m s$^{-1}$]	&	[m s$^{-1}$]	&	[m s$^{-1}$]\\
	\hline																
CCF RVs	&		&	3.1	&	2.2	&	3.0\\
\rule{0pt}{3.0ex}\logRHK\ & (\logRHK)$^{\mathrm{2}}$	&	1.8	&	2.1	&	2.9\\
\rule{0pt}{2.0ex}CCF FWHM & (CCF FWHM)$^{\mathrm{2}}$	&	1.6	&	2.0	&	2.4\\
\rule{0pt}{2.0ex}CCF BIS & (CCF BIS)$^{\mathrm{2}}$	&	2.4	&	2.1	&	2.9\\
\rule{0pt}{3.0ex}\logRHK\ & CCF FWHM	&	1.6	&	1.8	&	2.3\\
\rule{0pt}{2.0ex}\logRHK\ & CCF BIS	&	1.8	&	2.1	&	2.9\\
\rule{0pt}{2.0ex}CCF FWHM & CCF BIS	&	1.6	&	1.9	&	2.5\\
\rule{0pt}{3.0ex}SRA$_{\mathrm{F.}}$ shift & SRA$_{\mathrm{F.}}$ amp	&	1.1	&	0.9	&	1.0\\
    \hline
    \hline
    \end{tabular} \\
    \caption{Comparison of the linear RV modelling using different activity indicators for the three seasons of $\alpha$~Cen~B presented in Figure~\ref{fig:RVmodels}. \logRHK, the CCF FWHM and the CCF BIS were combined using Equation~\ref{eq:RVmodel} and the resulting RMS are compared to the one of the CCF RVs and the one obtained using the $\mathrm{RV_{SRA}}$ (combination of SRA$_{\mathrm{F.}}$ shift \& SRA$_{\mathrm{F.}}$ amp). While the choice of the activity indicators can lead to different RMS values, none of the combination display above can reach the RMS value obtained using the $\mathrm{RV_{SRA}}$.}
    \label{tab:2}
\end{table*}

\section{SRA shift variations}
\begin{figure*}
    \centering
    \includegraphics[width=1\textwidth]{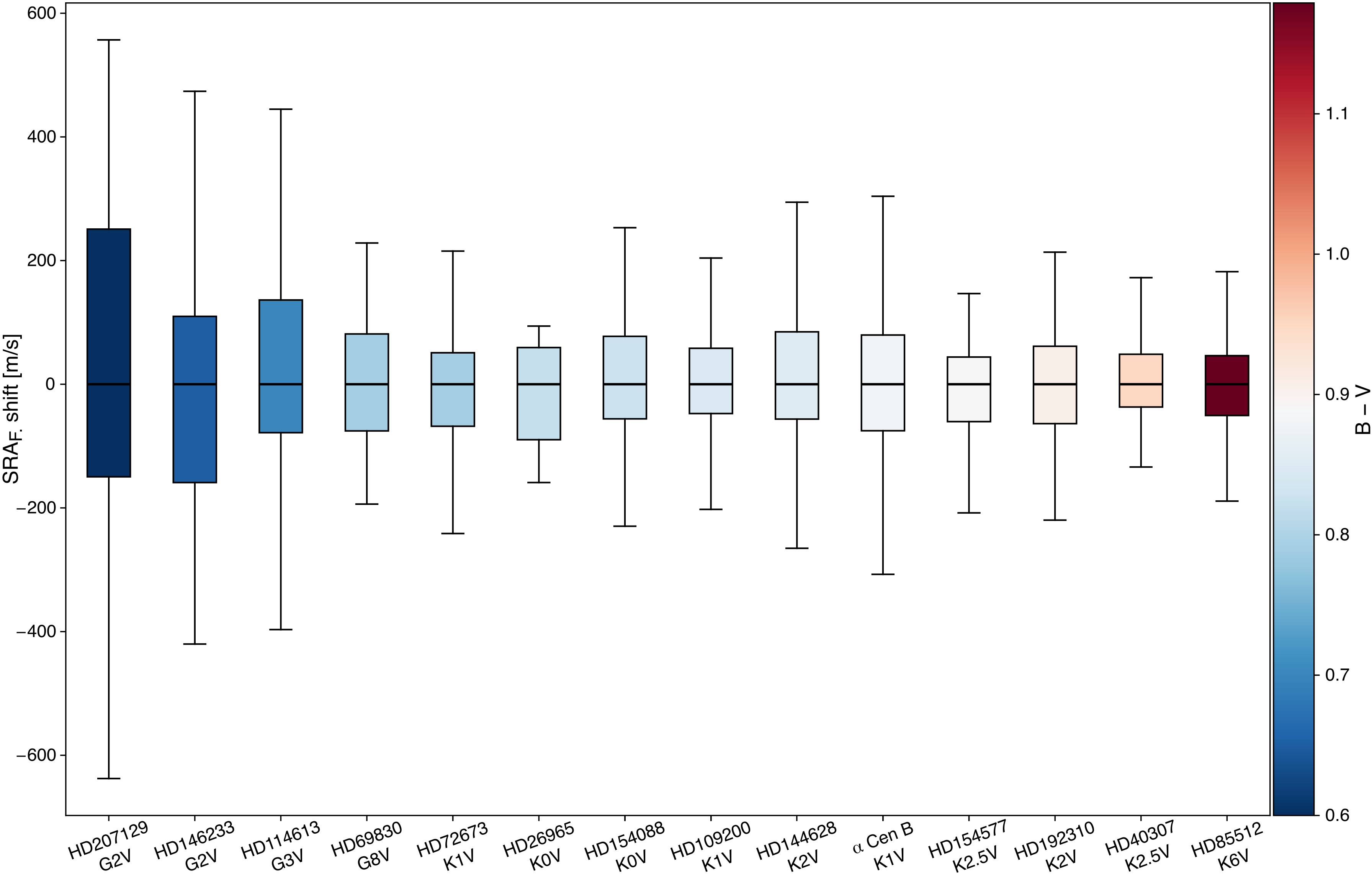}
    \caption{\label{fig:SRAvar} Measurements of the SRA$_{\mathrm{Feature}}$ shift variations for all \nstar~stars in our sample, ordered and colour coded using the $B - V$ colour index. The ordering of the stars follows the same order as Figure~\ref{fig:StarsFlux} and Table~\ref{tab:1} for ease of comparison. For each star, the bottom of the box, the black horizontal line and the top of the box correspond to the 25$^{\mathrm{th}}$, 50$^{\mathrm{th}}$, and 75$^{\mathrm{th}}$ percentile, respectively. The dispersion is larger for the hottest stars, as the SRA$_{\mathrm{Feature}}$ shifts are mainly influenced by the rotation of the star, which is generally faster for earlier type stars, resulting in larger variations.}
\end{figure*}


\section{Comparing the measured RMS using different stellar activity indicators for all \nstar~stars in our sample}
\begin{table*}
    \centering
    \begin{tabular}{l|c|c|c|c|c|c|c|c|c|c|c|c|c|c}
    \hline
    \hline
	&	CCF RVs	&	\logRHK\	&	CCF FWHM	&	CCF BIS	&	\logRHK\	&	\logRHK\	&	CCF FWHM	&	SRA$_{\mathrm{F.}}$ shift	\\
	&	[m s$^{-1}$]	&	(\logRHK)$^{\mathrm{2}}$	&	(CCF FWHM)$^{\mathrm{2}}$	&	(CCF BIS)$^{\mathrm{2}}$	&	CCF FWHM	&	CCF BIS	&	CCF BIS	&	SRA$_{\mathrm{F.}}$ amp	\\
	Stars&		&	[m s$^{-1}$]	&	[m s$^{-1}$]	&	[m s$^{-1}$]	&	[m s$^{-1}$]	&	[m s$^{-1}$]	&	[m s$^{-1}$]	&	[m s$^{-1}$]	\\
\hline																	
HD207129	&	3.7	&	1.9	&	2.5	&	2.2	&	2.0	&	2.0	&	2.2	&	1.9	\\
HD146233	&	2.9	&	2.6	&	2.7	&	2.7	&	2.7	&	2.6	&	2.7	&	2.4	\\
HD114613	&	3.8	&	2.6	&	2.7	&	2.7	&	2.5	&	2.6	&	2.5	&	2.3	\\
HD69830	&	1.8	&	1.4	&	1.3	&	1.5	&	1.3	&	1.4	&	1.4	&	1.2	\\
HD72673	&	1.5	&	1.2	&	1.1	&	1.2	&	1.1	&	1.1	&	1.1	&	0.8	\\
HD26965	&	0.9	&	0.7	&	0.6	&	0.6	&	0.7	&	0.7	&	0.6	&	0.5	\\
HD154088	&	1.7	&	1.6	&	1.5	&	1.5	&	1.5	&	1.5	&	1.5	&	1.3	\\
HD109200	&	1.6	&	1.4	&	1.2	&	1.5	&	1.3	&	1.3	&	1.2	&	1.2	\\
HD144628	&	1.7	&	1.5	&	\textcolor{red}{1.4}	&	1.6	&	1.5	&	1.5	&	1.5	&	1.5	\\
$\alpha$~Cen~B	&	4.1	&	3.1	&	1.9	&	2.6	&	1.9	&	2.9	&	2.3	&	1.4	\\
HD154577	&	1.3	&	1.2	&	1.1	&	1.2	&	1.0	&	1.2	&	1.1	&	1.0	\\
HD192310	&	2.0	&	1.5	&	1.5	&	1.5	&	1.4	&	1.5	&	1.4	&	1.4	\\
HD40307	&	1.8	&	1.1	&	1.1	&	1.2	&	1.0	&	1.1	&	1.1	&	1.0	\\
HD85512	&	1.0	&	0.9	&	0.9	&	0.9	&	\textcolor{red}{0.8}	&	0.9	&	0.9	&	0.9	\\
    \hline
    \hline
    \end{tabular} \\
    \caption{Comparison of the RMS obtained (in m s$^{-1}$) after combining different activity indicators using Equation~\ref{eq:RVmodel} for all \nstar~stars in our sample. Column 1 displays the RMS of the CCF RVs before any astrophysical noise mitigation. Each column after this presents the results after astrophysical noise mitigation using different combinations of parameters in Equation~\ref{eq:RVmodel}. The ordering of the stars follows the same order as Table~\ref{tab:1}. The values reported in the second and last column correspond to the results shown in the bottom panel of Figure~\ref{fig:RVmodels}. The activity indicator combinations used in the linear regression are the same as those presented in Table~\ref{tab:2}. As can be seen, while the choice of activity indicators leads to different RMS values, the results obtained using the $\mathrm{RV_{SRA}}$ (see the last column) show the lowest RMS values for the majority of the stars. In only two cases (highlighted in red), were we able to reach a RMS value slightly below the one obtained using $\mathrm{RV_{SRA}}$.}
    \label{tab:3}
\end{table*}


\section{Comparing the measured importances using different stellar activity indicators with random forest regression for all \nstar~stars in our sample}
\begin{table*}
    \centering
    \begin{tabular}{l ccc ccc ccc}
    \hline
    \hline
    & \multicolumn{3}{c}{Importances} & \multicolumn{3}{c}{Importances} & \multicolumn{3}{c}{Importances} \\
    \cmidrule(lr){2-4} \cmidrule(lr){5-7} \cmidrule(lr){8-10}
	Stars&	\logRHK\	&	SRA$_{\mathrm{F.}}$amp	&	SRA$_{\mathrm{F.}}$shift	&	CCF FWHM	&	SRA$_{\mathrm{F.}}$amp	&	SRA$_{\mathrm{F.}}$shift	&	CCF BIS	&	SRA$_{\mathrm{F.}}$amp	&	SRA$_{\mathrm{F.}}$shift	\\
\hline																		
HD207129	&	0.44	&	0.45	&	0.11	&	0.20	&	0.65	&	0.15	&	0.35	&	0.52	&	0.14	\\
HD146233	&	\textcolor{red}{0.43}	&	0.15	&	0.42	&	0.31	&	0.21	&	0.48	&	0.33	&	0.21	&	0.46	\\
HD114613	&	0.34	&	0.38	&	0.28	&	0.28	&	0.45	&	0.28	&	0.25	&	0.45	&	0.30	\\
HD69830	&	0.25	&	0.26	&	0.49	&	0.35	&	0.24	&	0.41	&	0.18	&	0.29	&	0.53	\\
HD72673	&	0.13	&	0.34	&	0.54	&	0.18	&	0.32	&	0.50	&	0.07	&	0.37	&	0.56	\\
HD26965	&	0.24	&	0.20	&	0.55	&	0.20	&	0.21	&	0.59	&	0.26	&	0.19	&	0.55	\\
HD154088	&	0.25	&	0.26	&	0.49	&	0.30	&	0.24	&	0.46	&	0.33	&	0.22	&	0.45	\\
HD109200	&	0.30	&	0.23	&	0.47	&	0.34	&	0.20	&	0.45	&	0.20	&	0.29	&	0.52	\\
HD144628	&	0.37	&	0.42	&	0.20	&	\textcolor{red}{0.52}	&	0.32	&	0.16	&	0.25	&	0.53	&	0.22	\\
$\alpha$~Cen~B	&	0.08	&	0.25	&	0.67	&	0.14	&	0.22	&	0.64	&	0.09	&	0.25	&	0.66	\\
HD154577	&	0.25	&	0.40	&	0.36	&	0.34	&	0.34	&	0.32	&	0.24	&	0.41	&	0.35	\\
HD192310	&	\textcolor{red}{0.39}	&	0.34	&	0.28	&	\textcolor{red}{0.44}	&	0.25	&	0.31	&	0.32	&	0.36	&	0.31	\\
HD40307	&	0.32	&	0.27	&	0.41	&	\textcolor{red}{0.38}	&	0.25	&	0.37	&	0.27	&	0.32	&	0.41	\\
HD85512	&	0.33	&	0.31	&	0.36	&	\textcolor{red}{0.39}	&	0.28	&	0.33	&	0.28	&	0.34	&	0.38	\\
    \hline
    \hline
    \end{tabular} \\
    \caption{Comparison of the importances for different activity indicator combinations using random forest regression for the \nstar~stars in our sample. The ordering of the stars follows the same as that of Table~\ref{tab:1} (i.e. ordered by $B - V$). The values reported in the first three columns correspond to the results shown in the bottom panel of Figure~\ref{fig:RVmodel_RF}. While the choice of activity indicator can lead to different results, the SRA$_{\mathrm{Feature}}$ shifts show the highest importance in most cases. We have highlighted in red where the importance is greater for \logRHK, CCF FWHM or CCF BIS, which corresponds primarily to targets with lower SNR and/or poor sampling.}
    \label{tab:4}
\end{table*}


\bsp	
\label{lastpage}
\end{document}